\documentclass[12pt]{article}
\usepackage{a4wide,epsfig,psfrag,amsmath,amssymb,cite,scalefnt}
\usepackage{color}

\graphicspath{ {figs/} }

\newcommand{\be}{\begin{equation}}
\newcommand{\ee}{\end{equation}}
\newcommand{\bea}{\begin{eqnarray}}
\newcommand{\eea}{\end{eqnarray}}

\def\LO{\ifmmode \textrm{LO} \else LO\fi}
\def\NLO{\ifmmode \textrm{NLO} \else NLO\fi}
\def\NNLO{\ifmmode \textrm{NNLO} \else NNLO\fi}
\newcommand{\sof}{\frac{\sigma^{(1)}_\textrm{fin}}{\sigma^{\rm LO}}}
\newcommand{\stf}{\frac{\sigma^{(2)}_\textrm{fin}}{\sigma^{\rm LO}}}
\newcommand\numberthis{\addtocounter{equation}{1}\tag{\theequation}}

\allowdisplaybreaks

\newcommand{\gsim}{\;\rlap{\lower 3.5 pt \hbox{$\mathchar \sim$}} \raise 1pt
 \hbox {$>$}\;}
\newcommand{\lsim}{\;\rlap{\lower 3.5 pt \hbox{$\mathchar \sim$}} \raise 1pt
 \hbox {$<$}\;}

\begin{document}

\title{\vskip-3cm{\baselineskip14pt
    \begin{flushleft}
      \normalsize DESY 15-135, TTP15-026
  \end{flushleft}}
  \vskip1.5cm
  Higgs boson pair production: top quark mass effects at 
  NLO and NNLO
}

\author{
  Jonathan Grigo$^{(a)}$, 
  Jens Hoff$^{(b)}$,
  Matthias Steinhauser$^{(a)}$
  \\[1em]
  {\small\it (a) Institut f{\"u}r Theoretische Teilchenphysik,
  Karlsruhe Institute of Technology (KIT)}\\
  {\small\it 76128 Karlsruhe, Germany}
  \\[1em]
  {\small\it (b) Deutsches Elektronen Synchrotron DESY, Platanenallee 6, 15738
  Zeuthen, Germany}
}

\date{}

\maketitle

\thispagestyle{empty}

\begin{abstract}
  We compute next-to-next-to-leading order QCD corrections to the
  gluon-induced production cross section of Higgs boson pairs in the large top
  quark mass limit using the soft-virtual approximation. In the limit of
  infinitely-heavy top quark we confirm the results in the literature. We add
  two more expansion terms in the inverse top quark mass to the $M_t\to\infty$
  result.  Since the $1/M_t$ expansion converges poorly, we try to improve on
  it by factorizing the exact leading order cross section. We discuss two ways of
  doing that and conclude that the finite top quark mass effects shift the
  cross section at most by about 10\% at next-to-leading order and
  by about 5\% at next-to-next-to-leading order.

  \medskip


\end{abstract}

\thispagestyle{empty}

%- }}}

\newpage

%- {{{ Introduction:

\section{Introduction}

In the coming years, one of the main tasks in particle physics is the
understanding of the mechanism of the electroweak symmetry breaking. 
After the experimental determination of the Higgs boson mass,
the Higgs potential is fully fixed in the Standard Model. However,
it is very important to independently measure the 
self-coupling of the Higgs boson, which can be obtained from studying the
production of Higgs boson pairs. Since the corresponding cross section is
${\cal O}(10^3)$ times smaller than the one for single Higgs
boson production, Higgs boson pair production poses a challenging problem
for the LHC, even after the luminosity upgrade around 2020.

There are a number of phenomenological analyses which investigate the
possibility to extract the self coupling from cross section
measurements.  First studies have been performed more than 15 years
ago~\cite{Djouadi:1999rca,Baur:2002qd,Baur:2003gp}. Since the discovery of the
Higgs boson there has been an increasing interest in this topic and a number
of refined analyses have been performed, see, e.g.,
Refs.~\cite{Dolan:2012rv,Papaefstathiou:2012qe,Baglio:2012np,Goertz:2013kp,Gouzevitch:2013qca}.

Higgs boson pairs can be produced via the fusion of two partons or vector
bosons, via the radiation off vector bosons, or in
association with heavy quarks. Similar to single Higgs boson production, the
numerically dominant mechanism is gluon fusion although the leading order (LO)
contribution is loop-suppressed.  Due to the larger Yukawa coupling, 
the dominant contribution comes from top quark loops in the
Standard Model.  For this
reason we concentrate in this paper on such contributions.

For the LO order corrections the exact dependence on the top quark mass and
the center-of-mass energy is known~\cite{Glover:1987nx,Plehn:1996wb}.  At
next-to-leading (NLO) QCD corrections have been computed for the first time
more than 15 years ago~\cite{Dawson:1998py,Grigo:2013rya} in
the infinite top quark mass limit using an
effective theory.  Finite top quark mass effects have been investigated in
Ref.~\cite{Grigo:2013rya} where a systematic expansion in the inverse top
quark mass has been applied and a quantitative estimate of the quark mass
effects has been provided. It has been estimated that they do not exceed
${\cal O}(10\%)$ of the NLO contribution. Finite top quark mass 
effects have also been considered in
Ref.~\cite{Maltoni:2014eza} where the exact real radiation
contribution is combined with the effective-theory virtual corrections. As a
result, a reduction of about $-10\%$ of the cross section is obtained. We will
comment in Section~\ref{sec::nlo} on this issue.

Within the effective theory also next-to-next-to-leading (NNLO) contributions
are available~\cite{deFlorian:2013uza,deFlorian:2013jea}. In this context it
is interesting to note that the three-loop matching coefficient of the
effective operator for two Higgs bosons and two, three or four gluons is
different from the one for single Higgs boson
production~\cite{Grigo:2014jma}. The results for the virtual corrections
obtained in Ref.~\cite{deFlorian:2013uza} have been cross-checked in
Ref.~\cite{Grigo:2014jma} where the calculation has been performed 
without reference to the effective theory.  The
resummation of threshold-enhanced logarithms to next-to-next-to-leading
logarithmic (NNLL) accuracy has been performed in
Refs.~\cite{Shao:2013bz,deFlorian:2015moa}.

The remainder of this paper is organized as follows: In the next Section we
review the construction of the soft-virtual approximation for the production
cross section and discuss two options to factorize the exact LO result from
the higher order contributions. We argue that a factorization at the level of
the differential cross section w.r.t. the Higgs boson pair invariant
mass leads to more stable results. Afterwards we reconsider in
Section~\ref{sec::nlo} the top mass corrections at NLO.  Virtual NNLO
corrections including finite top quark mass effects are computed in
Section~\ref{sec::nnlo}. They are used in Section~\ref{sec::impr_nnlo} to
present phenomenological results for Higgs pair production up to
NNLO. Section~\ref{sec::con} contains our conclusions.

%- }}}
%- {{{ Factorizing the exact LO expression:

\section{\label{sec::fac}Factorizing the exact LO expression}

We write the perturbative expansion of the partonic cross section
for the production of Higgs boson pairs in the form
\begin{eqnarray}
  \sigma_{ij \to HH + X}(s,\rho) = \delta_{ig} \delta_{jg}
  \sigma_{gg}^{(0)}(s,\rho) 
  + \frac{\alpha_s}{\pi} \sigma^{(1)}_{ij}(s,\rho)
  + \left(\frac{\alpha_s}{\pi}\right)^2 \sigma^{(2)}_{ij}(s,\rho)
  + \ldots
  \,,
\end{eqnarray}
where $\alpha_s\equiv\alpha_s^{(5)}(\mu)$, 
$\sqrt{s}$ is the partonic center-of-mass energy and 
$ij\in\{gg,qg,\bar{q}g,q\bar{q}\}$. 
Since the quark-induced channels are
numerically small~\cite{Dawson:1998py} we consider in this paper  
only the $gg$ channel. We use the variable
\begin{eqnarray}
  \rho &=& \frac{m_H^2}{M_t^2}\,,
\end{eqnarray}
to parametrize the dependence of the cross section
on the Higgs boson and top quark mass.
We renormalize the top quark mass in the on-shell
scheme. Furthermore, we set the factorization and renormalization scale
equal to each other and write $\mu=\mu_r=\mu_f$.

For later convenience we introduce for $\sigma_{gg \to HH + X}$
\begin{eqnarray}
  \sigma_{gg \to HH + X}(s,\rho) &=&
  \sigma^{\rm LO} + \delta\sigma^{\rm NLO} + \delta\sigma^{\rm NNLO} + \ldots 
  \,,
\end{eqnarray}
and denote the sum of the first two terms on the 
right-hand side by $\sigma^{\rm NLO} = \sigma^{\rm LO} + \delta\sigma^{\rm
  NLO}$. 

Finite top quark mass effects to $gg\to HH$ at NLO have been considered for
the first time in Ref.~\cite{Grigo:2013rya}. The applied method is based on
``reversed unitarity''~\cite{Anastasiou:2002yz} where, with the help of the
optical theorem, the imaginary part of forward scattering amplitudes are
computed to obtain the total cross section. The virtual corrections have also
been computed by directly considering the $gg\to HH$ amplitude.  Expansion
terms up to order $1/M_t^{12}$ of the NLO contribution to $\sigma(pp\to HH)$
have been computed~\cite{Grigo:2013rya,Grigo:2013xya,Grigo:2014oqa}.  The
factorization of the exact LO corrections has then been implemented at the
partonic level for the total cross section using
\begin{eqnarray}
  \label{eq::nlo_fact1}
  \sigma_{gg,N}^{(1)} &=& 
  \sigma_{gg,\rm exact}^{(0)}  \Delta_{gg}^{(N)},
  \;\;\;\;\;  \Delta_{gg}^{(N)} = 
  \frac{\sigma_{gg,\rm exp}^{(1)}}{\sigma_{gg,\rm exp}^{(0)}} = 
  \frac{\sum \limits_{n=0}^{N} c_{gg,n}^{\rm NLO}\rho^n}{\sum
    \limits_{n=0}^{N} c_{gg,n}^{\rm LO}\rho^n}
  \,,
\end{eqnarray}
where $\sigma_{gg,\rm exact}^{(0)}$ contains the exact dependence on $s$ and
$\rho$.

In Ref.~\cite{Dawson:1998py}, where the NLO result has been computed for the
first time in the heavy top quark limit, a different approach has been
applied. Actually, the exact LO result has been factorized before
the integration over the Higgs pair invariant mass. In this approach the
(total) NLO cross section reads
\begin{eqnarray}
  \sigma_{gg}^{(1)} &=& 
  \int_{4m_H^2}^{s} {\rm d}Q^2 \,
  \frac{ {\rm d}\sigma^{(0)}_{gg,\rm exact} } { {\rm d}Q^2 } 
  \,
  \frac{ \frac{ {\rm d}\sigma^{( 1 )}_{gg,\rm exp} } { {\rm d}Q^2 } }
  { \frac{ {\rm d}\sigma^{( 0 )}_{gg,\rm exp} } { {\rm d}Q^2 } }
  \,,
  \label{eq::nlo_fact2}
\end{eqnarray}
where 'exact' and 'exp' reminds whether exact or (in $\rho$) expanded results
are used. $Q^2$ is the invariant mass squared of the Higgs boson pair.
Expressed in terms of ${\rm d}\sigma/{\rm d}Q^2$ it is possible to re-write
Eq.~(\ref{eq::nlo_fact1}) as
\begin{eqnarray}
  \sigma_{gg}^{(1)} &=& \frac{\sigma_{gg,\rm exact}^{(0)}}{\sigma_{gg,\rm
      exp}^{(0)}} \, \int_{4m_H^2}^{s} {\rm d}Q^2 \, \frac{ {\rm
      d}\sigma^{(1)}_{gg,\rm exp} } { {\rm d}Q^2 } \,.
  \label{eq::nlo_fact3} 
\end{eqnarray}
From the comparison of Eqs.~(\ref{eq::nlo_fact2}) and~(\ref{eq::nlo_fact3})
one expects that~(\ref{eq::nlo_fact2}) leads to better results
since the differential factorization (DF) in Eq.~(\ref{eq::nlo_fact2}) 
results in a better-behaved
integrand, in particular for $Q^2>4M_t^2$.

For the virtual corrections, which are proportional to $\delta(s-Q^2)$, 
one has immediate access to the $Q^2$ dependence
and the DF of Eq.~(\ref{eq::nlo_fact2}) can be
applied.\footnote{In fact, for the virtual corrections
  Eqs.~(\ref{eq::nlo_fact1}) and~(\ref{eq::nlo_fact2}) are equivalent.}
The real corrections, however, are obtained with the
help of the optical theorem which directly leads to the total cross section
and thus Eq.~(\ref{eq::nlo_fact2}) can not be used.
For the construction of the soft-virtual (SV) approximation, which is
discussed below, we need in addition to the virtual term also the contributions
from soft gluon emission. Since the soft contributions are proportional to 
the LO Born cross section Eq.~(\ref{eq::nlo_fact2}) can immediately be applied
to the SV cross section.

In the following we discuss the construction of the SV approximation for the
cross section (see also Ref.~\cite{deFlorian:2012za}). For simplicity we
consider for the following schematic reasoning the total cross section
$\sigma$. Note, however, that the arguments also hold for ${\rm d}\sigma /
{\rm d} Q^2$.  In a first step we split $\sigma$ according to
\begin{eqnarray}
  \sigma &=& \sigma^{\rm virt + ren} + \sigma^{\rm real + split}
  \,,
    \label{eq::sig_split}
\end{eqnarray}
where the two terms on the right-hand side correspond to the virtual
corrections (including ultra-violet counterterms) and 
the real corrections (including the contributions from the factorization 
of initial-state singularities).
They are individually divergent and only the sum is
finite. In the next step we re-organize the terms on the right-hand side such
that $\sigma$ can be written as
\begin{eqnarray} 
  \sigma &=& \Sigma_{\rm SV} + \Sigma_{\rm H}
  \label{eq::sig_SV_H}
  \,,
\end{eqnarray}
where ``SV'' refers to ``soft-virtual'' and ``H'' to ``hard''.  This is
achieved by splitting $\sigma^{\rm virt + ren}$ into a divergent and a finite
term and separating $\sigma^{\rm real + split}$ into a (divergent) soft and
(finite) hard contribution. The soft contribution is combined with
$\sigma^{\rm virt + ren}$ to obtain $\Sigma_{\rm SV}$ such that
\begin{eqnarray}
  \Sigma_{\rm SV}
  &=& \Sigma_{\rm div} + \Sigma_{\rm fin} 
  + \Sigma_{\rm soft}^{\rm real + split}
  \,, \nonumber\\
  \Sigma_{\rm H}
  &=& \Sigma_{\rm hard}^{\rm real + split}
  \,.
  \label{eq::sig_SV_H_2}
\end{eqnarray}
$\Sigma_{\rm SV}$ and $\Sigma_{\rm H}$ are separately finite.
$\Sigma_{\rm div}$ is constructed following Ref.~\cite{Catani:1998bh};
explicit expressions can be found in
Refs.~\cite{deFlorian:2012za,Grigo:2014jma}.
Note that the finite part is constructed solving the equation
\begin{eqnarray}
  \sigma^{\rm virt + ren} = \Sigma_{\rm fin} + \Sigma_{\rm div}
  \,,
  \label{eq::sig_fin}
\end{eqnarray}
for $\Sigma_{\rm fin}$.
In this paper $\sigma^{\rm virt + ren}$ is computed including top quark mass
effects. Mass effects are automatically taken into account in $\Sigma_{\rm div}$ and
$\Sigma_{\rm soft}^{\rm real + split}$ since these contributions are
proportional to the exact LO cross section~\cite{deFlorian:2012za}.
Note that
Eqs.~(\ref{eq::sig_split}),~(\ref{eq::sig_SV_H}),~(\ref{eq::sig_SV_H_2})
and~(\ref{eq::sig_fin}) also hold for the differential cross section ${\rm
  d}\sigma/{\rm d}Q^2$ and thus a factorization as suggested in
Eq.~(\ref{eq::nlo_fact2}) can be performed for the soft-virtual contribution.

To be more specific we present the NLO and NNLO
differential version of Eq.~(\ref{eq::sig_fin}) which
reads~\cite{deFlorian:2012za,Grigo:2014jma} 
\begin{align} 
  & \frac{{\rm d}\sigma_v^{(1)}}{{\rm d}t}
  = \frac{{\rm d}\sigma_{v,\rm fin}^{(1)}}{{\rm d}t}
  + 2 \mbox{Re}\left[I_g^{(1)}\right] \frac{{\rm d}\sigma^{(0)}}{{\rm d}t}
  \,,
  \nonumber\\
  & \frac{{\rm d}\sigma_v^{(2)}}{{\rm d}t}
  = \frac{{\rm d}\sigma_{v,\rm fin}^{(2)}}{{\rm d}t}
  + 2 \mbox{Re}\left[I_g^{(1)}\right] \frac{{\rm d}\sigma_{v,\rm fin}^{(1)}}{{\rm d}t}
  + \left\{ \left|I_g^{(1)}\right|^2 
  + 2 \mbox{Re}\left[\left(I_g^{(1)}\right)^2\right]
  + 2 \mbox{Re}\left[I_g^{(2)}\right]
  \right\}
  \frac{{\rm d}\sigma^{(0)}}{{\rm d}t}
  \,.
  \label{eq::sig_fin_2}
\end{align}
where $t=(q_1-q_3)^2$ with $q_1$ ($q_3$) being the incoming (outgoing) momentum of a
gluon (Higgs boson), $\sigma^{(0)}\equiv\sigma^{\rm LO}$ and
explicit expressions for the operators $I_g^{(1,2)}$
can be found in Ref.~\cite{Catani:1998bh}.

We adopt the notation from Ref.~\cite{deFlorian:2012za} and
parametrize radiative corrections to the partonic cross section via
\begin{eqnarray}
  Q^2 \frac{ {\rm d}\sigma }{ {\rm d} Q^2 } &=& \sigma^{\rm LO} z G(z)
  \,,
  \label{eq::G}
\end{eqnarray}
with 
\begin{eqnarray}
  z &=& \frac{Q^2}{s}
  \,,
\end{eqnarray}
and
\begin{eqnarray}
  G(z) &=& \delta(1-z) + \frac{\alpha_s}{2\pi} G^{(1)}(z)
  + \left(\frac{\alpha_s}{2\pi}\right)^2 G^{(2)}(z) + \ldots
  \nonumber\\
  &=& G_{\rm SV}(z) + G_{\rm H}(z)
  \,,
  \label{eq::G_2}
\end{eqnarray}
where the renormalization scale dependence in the strong coupling
constant $\alpha_s$ is suppressed. 
From Eq.~(\ref{eq::G})
one obtains for the total cross section
\begin{eqnarray}
  \sigma &=& \int_{4 m_H^2}^s {\rm d}Q^2 
  \frac{ {\rm d}\sigma }{ {\rm d} Q^2 }
  \,\,=\,\, \int_{1-\delta}^1 {\rm d}z \, \sigma^{\rm LO}(zs) \, G(z)
  \,,
  \label{eq::sigma_tot}
\end{eqnarray}
with 
\begin{eqnarray}
  \delta &=& 1-\frac{4m_H^2}{s}\,.
  \label{eq::delta}
\end{eqnarray}
In the second line of Eq.~(\ref{eq::G_2}) we split $G(z)$ into soft-virtual
and hard contribution. Note that in our approach we do not have access to
$G_{\rm H}(z)$. Actually, at NNLO we only have $G_{\rm SV}(z)$ at hand and at 
NLO only the heavy top expansion of
$\int_{1-\delta}^1 {\rm d}z \, \sigma^{\rm LO}(zs) \, G_{\rm H}(z)$
is available to us.

Explicit results for $G_{\rm SV}(z)$ can be found in
Ref.~\cite{deFlorian:2012za} including higher order terms in $\epsilon$
specifying, however, the renormalization and factorization scale to
$\sqrt{s}\simeq\sqrt{Q^2}$.
For completeness we provide the NLO and NNLO
results for $G_{\rm SV}(z)$ for generic $\mu$ in the limit $\epsilon\to
0$. The results read
\begin{align}
G_\textrm{SV}^{(1)} &= D_{-1} \Big[\frac{2\pi^2}{3}C_A + \sof\Big] - 4C_ALD_0 + 8C_AD_1  
\,,
\nonumber\\
G_\textrm{SV}^{(2)} &= 
 D_{-1} \Bigg\{-\frac{4\pi^2}{3}C_A^2L^2 + \frac{11\pi^2}{108}n_l^2 + \stf + 
   L\Bigg[-\frac{1}{3}\sof n_l \nonumber\\
& \quad + C_A\left(\left(-\frac{1}{3} - \frac{2\pi^2}{9}\right) n_l + \frac{11}{6} \sof  \right)+ 
     C_A^2\left(\frac13 + \frac{11\pi^2}{9} - 38\zeta_3\right)\Bigg] \nonumber\\
& \quad + C_A^2\left(\frac{607}{81} + \frac{517\pi^2}{108} - \frac{\pi^4}{80} - \frac{407\zeta_3}{36}\right) + C_A\left[\frac{2\pi^2}{3}\sof + n_l\left(-\frac{82}{81} - \frac{11\pi^2}{8} +\frac{37}{18}\zeta_3\right)\right]\Bigg\}\nonumber\\
& \quad + D_0 \Bigg\{L^2\left(-\frac{11}{3}C_A^2 + \frac{2}{3}C_An_l\right) + C_An_l \left(\frac{56}{27} - \frac{4\pi^2}{9}\right) + 
   L\Bigg[C_A^2 \left(-\frac{134}{9} + \frac{10\pi^2}{3}\right) \nonumber\\
& \quad + C_A \left(\frac{20}{9}n_l - 4\sof\right)\Bigg] + 
   C_A^2\left(-\frac{404}{27} + \frac{22\pi^2}{9} + 78\zeta_3\right) \Bigg\} + D_1 \Bigg[16C_A^2L^2\nonumber\\
& \quad + L\left(\frac{44}{3}C_A^2 - \frac{8}{3} C_An_l\right) + C_A^2\left(\frac{268}{9} - \frac{20\pi^2}{3}\right) + C_A\left(-\frac{40}{9}n_l + 8\sof\right)\Bigg] \nonumber\\
& \quad + D_2 \Bigg[-\frac{44}{3}C_A^2 - 48C_A^2L + \frac{8}{3}C_An_l\Bigg] + 
 32C_A^2D_3
 \,,
\label{eq::G_SV}
\end{align}
where $C_F=4/3, C_A=3$ are QCD colour factors, $n_l=5$ is the number of
massless quarks, $L= \log (\mu^2/s)$ and
\begin{align}
  D_{-1} &= \delta(1-z) \ , & D_{n \geq 0} = \left[\frac{\log^n
      (1-z)}{1-z}\right]_+ 
  \,.
\end{align}
The quantities $\sigma_{\rm fin}^{(1)}$ and $\sigma_{\rm fin}^{(2)}$ are
evaluated for $\mu=\sqrt{s}$. They are
obtained from ${\rm d}\sigma_{v,\rm fin}^{(1)} / {\rm d}t$ and ${\rm
  d}\sigma_{v,\rm fin}^{(2)} / {\rm d}t$ in Eq.~(\ref{eq::sig_fin_2})
after integration over $t$.

In the next Section we apply the formalism described in this Section
at NLO and compare to the results obtained in Ref.~\cite{Grigo:2013rya}.
The numerical effects at NNLO are presented in Section~\ref{sec::impr_nnlo}
using the mass corrections computed in Section~\ref{sec::nnlo}.

%- }}}
%- {{{ Revisiting NLO:

\section{\label{sec::nlo}Revisiting NLO}

In this Section we restrict ourselves to NLO
and compare the results of Ref.~\cite{Grigo:2013rya} with the
alternative factorization discussed in the previous section.

To obtain numerical results we employ MSTW2008 parton distribution functions
(PDFs)~\cite{Martin:2009iq} and consistently use N$^k$LO PDFs to compute N$^k$LO
($k=0,1,2$) cross sections.  We assume the energy of the LHC to be $14~{\rm
  TeV}$ and adopt the values of the strong coupling constants $\alpha_s(M_{\rm
  Z})$ that we use in our computation from the MSTW PDF fit:
\begin{align}
  \alpha_s^{\rm LO}  =0.13939\,,\quad
  \alpha_s^{\rm NLO} =0.12018\,,\quad
  \alpha_s^{\rm NNLO}=0.11707\,.
\end{align}
We renormalize the top quark in the on-shell scheme
and use $M_t=173.21$~GeV~\cite{Agashe:2014kda}. For the Higgs boson we use
$m_H=125.09$~GeV~\cite{Aad:2015zhl}.
For numerical results shown in this Section the renormalization and
factorization scales have been set to $2m_H$.

\begin{figure}[t]
  \begin{center}
    \includegraphics[width=1.0\columnwidth]{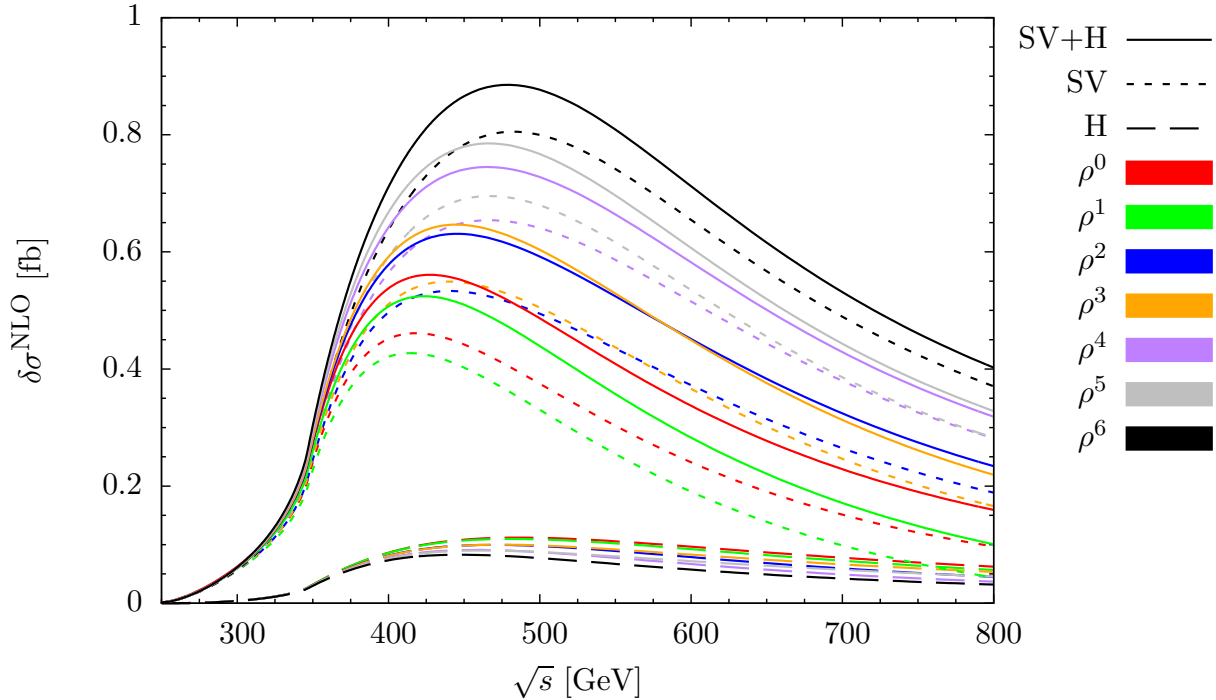}
    \\
    \caption{\label{fig::sig_part}
      Partonic cross section as a function the center-of-mass energy including
      various orders in the inverse top quark mass. The dotted and dashed
      curves show the breakup of the complete result (solid lines) into
      soft-virtual and hard contribution.}
  \end{center}
\end{figure}

In Fig.~\ref{fig::sig_part} we show the results for the partonic cross
section as a function of $\sqrt{s}$ from Fig.~6 of Ref.~\cite{Grigo:2013rya},
see solid lines. The splitting of these results into soft-virtual and hard
contributions is shown as dotted and dashed curves, respectively.  For the
complete result and soft-virtual approximation
the infinite-top quark result corresponds to the (red) second line
from below. The lowest line includes in addition the $1/M_t^2$
corrections. The curves including higher order mass corrections are above the
infinite-top quark result. The topmost curves include $1/M_t^{12}$ terms.
The hard contribution shows a quite flat behaviour above
$\sqrt{s}\gsim 400$~GeV and exhibits smaller shifts when including
top mass corrections. Furthermore, the infinite top mass result
corresponds to the topmost curve and the lowest curve includes $1/M_t^{12}$ terms.
From Fig.~\ref{fig::sig_part} it is evident that 
the hard contribution is numerically much smaller than the soft-virtual one.

\begin{figure}[t]
  \begin{center}
    \includegraphics[width=1.0\columnwidth]{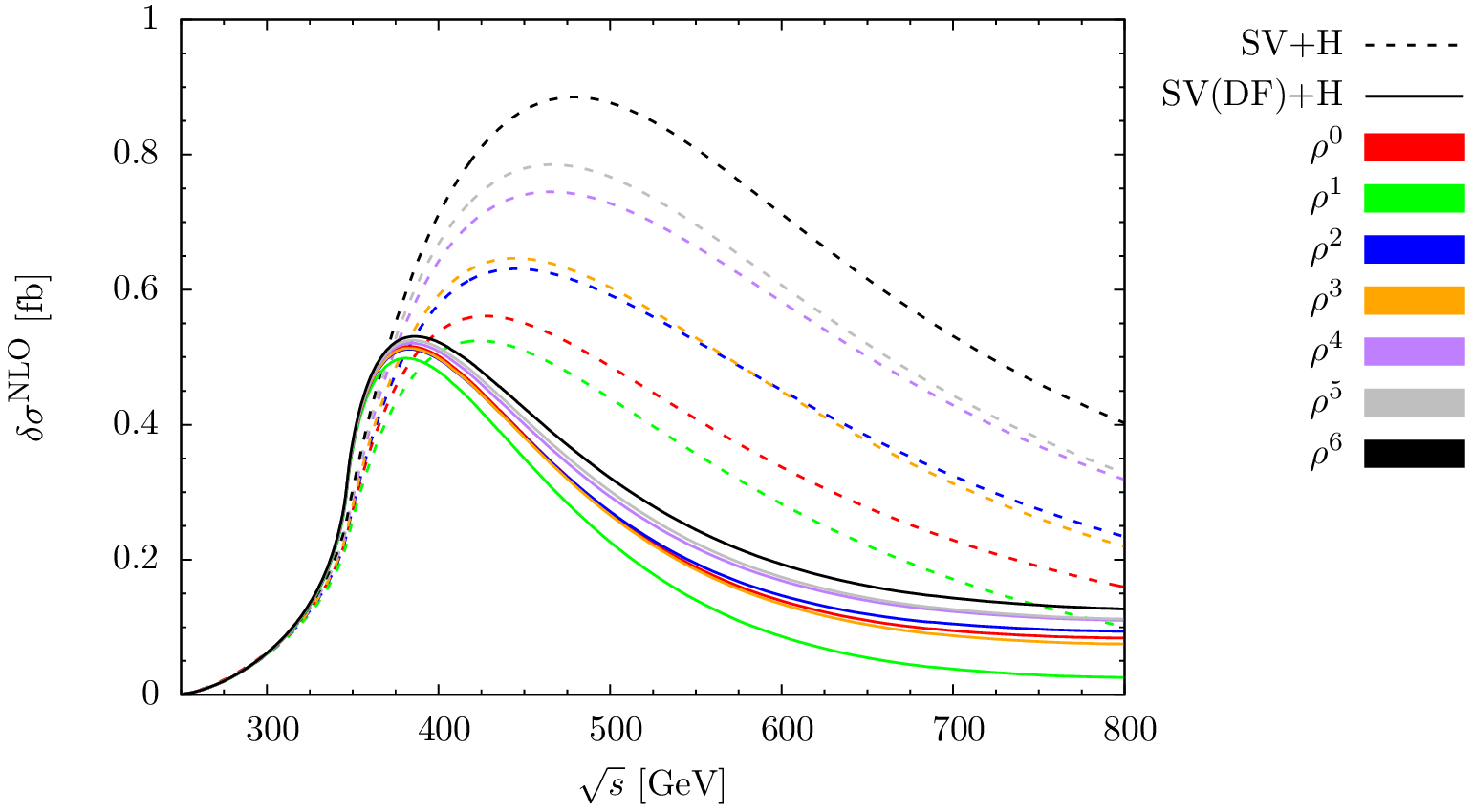}
    \\
    \caption{\label{fig::sig_part_2}
      Partonic cross section as a function the center-of-mass energy including
      various orders in the inverse top quark mass. The dashed and solid
      curves correspond to the factorization for the total and differential
      cross section, respectively. The colour coding is taken over from
      Fig.~\ref{fig::sig_part}.}
  \end{center}
\end{figure}

In Fig.~\ref{fig::sig_part_2} 
we compare the solid curves from
Fig.~\ref{fig::sig_part} (which are dotted in this plot)
with the results obtained with the help of DF applied to the SV approximation
and adding the hard contribution as given
by the dashed lines in Fig.~\ref{fig::sig_part}.
The relative position of the lines and the colour coding is as in
Fig.~\ref{fig::sig_part}.  
For lower values of $\sqrt{s}$ the two approaches lead to comparable results.
However, the DF curves have their maximum at lower values of $\sqrt{s}$ and
lead to a smaller cross section for larger values of $\sqrt{s}$. Furthermore,
one observes a drastic improvement in the convergence behaviour when including
higher order mass corrections. In particular, for $\sqrt{s}=400$~GeV 
the difference between the 
infinite top mass result and the one including $1/M_t^{12}$ terms
amounts to only $\approx0.05$~fb to be compared with $\approx0.25$~fb
for the dashed curves.

\begin{figure}[t]
  \begin{center}
    \includegraphics[width=1.0\columnwidth]{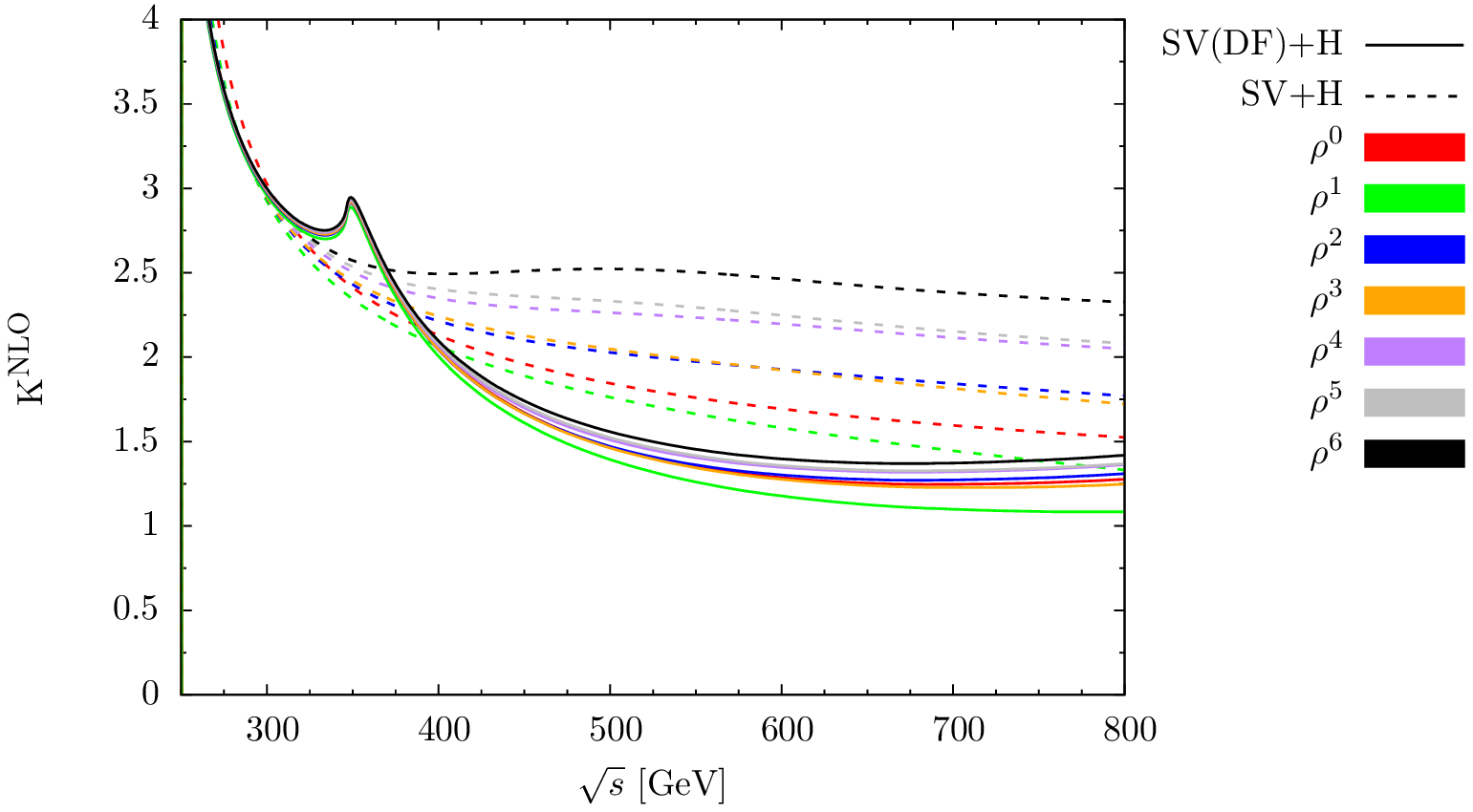}
    \\
    \caption{\label{fig::sig_K_part}
      Partonic NLO $K$ factor for the factorization performed at the 
      level of the total (dashed) and differential (solid) cross section.
    }
  \end{center}
\end{figure}

At this point we want to stress that the splitting between the
soft-virtual and hard contribution in Eq.~(\ref{eq::sig_SV_H}) is
arbitrary. In fact, the soft-virtual contribution of $G(z)$, $G_{\rm SV}(z)$,
is constructed for the limit $z\to1$ and thus it is possible to
replace $G_{\rm SV}(z)$ by $f(z)G_{\rm SV}(z)$ with $f(1)=1$ (see, e.g.,
discussions  in
Refs.~\cite{Anastasiou:2014vaa,Herzog:2014wja}). The hard contribution
is modified accordingly such that the sum of $\Sigma_{\rm SV} +
\Sigma_{\rm H}$ does not change.  One observes that for $f(z)=z$ the
soft-virtual contribution approximates the partonic NLO 
contribution to the cross
section very well\footnote{Note that the corresponding curves are not
shown in Fig.~\ref{fig::sig_part_2}.} such that at the hadronic level the
deviations are below $2\%$.
Based on this observation we use $f(z)=z$ for the NNLO cross section
where we only have the soft-virtual approximation at hand.
Furthermore, better results are obtained by replacing $L$ in
Eq.~(\ref{eq::G_SV}) by $L=\log (\mu^2/Q^2)$ which is justified since
$\sqrt{s}\approx\sqrt{Q^2}$ in the soft limit.

Note that in Ref.~\cite{deFlorian:2013jea} it has been observed that the soft-virtual
approximation constructed in Mellin space approximates the full
(effective-theory) result with an accuracy of 2\%.

It is interesting to look at the partonic $K$ factor which is defined via
\begin{align}
  K^\NLO &= \frac{\sigma^\LO + \delta\sigma^\NLO}{\sigma^\LO}
  \,.
  \label{eq::K_NLO}
\end{align}
Results for the two methods to factorize the exact LO term are
plotted in Fig.~\ref{fig::sig_K_part} as a function of $\sqrt{s}$ where the
dashed curves are already shown in Ref.~\cite{Grigo:2013rya}.  One observes
that DF leads to a lower $K$ factor and that the spread among the various
$\rho$ orders is smaller. Furthermore, it is interesting to note that for DF
the top quark pair threshold behaviour of the LO term is not washed out in
contrast to the dashed curves. It is
common to both factorization methods that there is a strong raise when
approaching the threshold for Higgs boson pair production (see also discussion
in Ref.~\cite{Grigo:2013rya}).

\begin{figure}[t]
  \begin{center}
    \includegraphics[width=1.0\columnwidth]{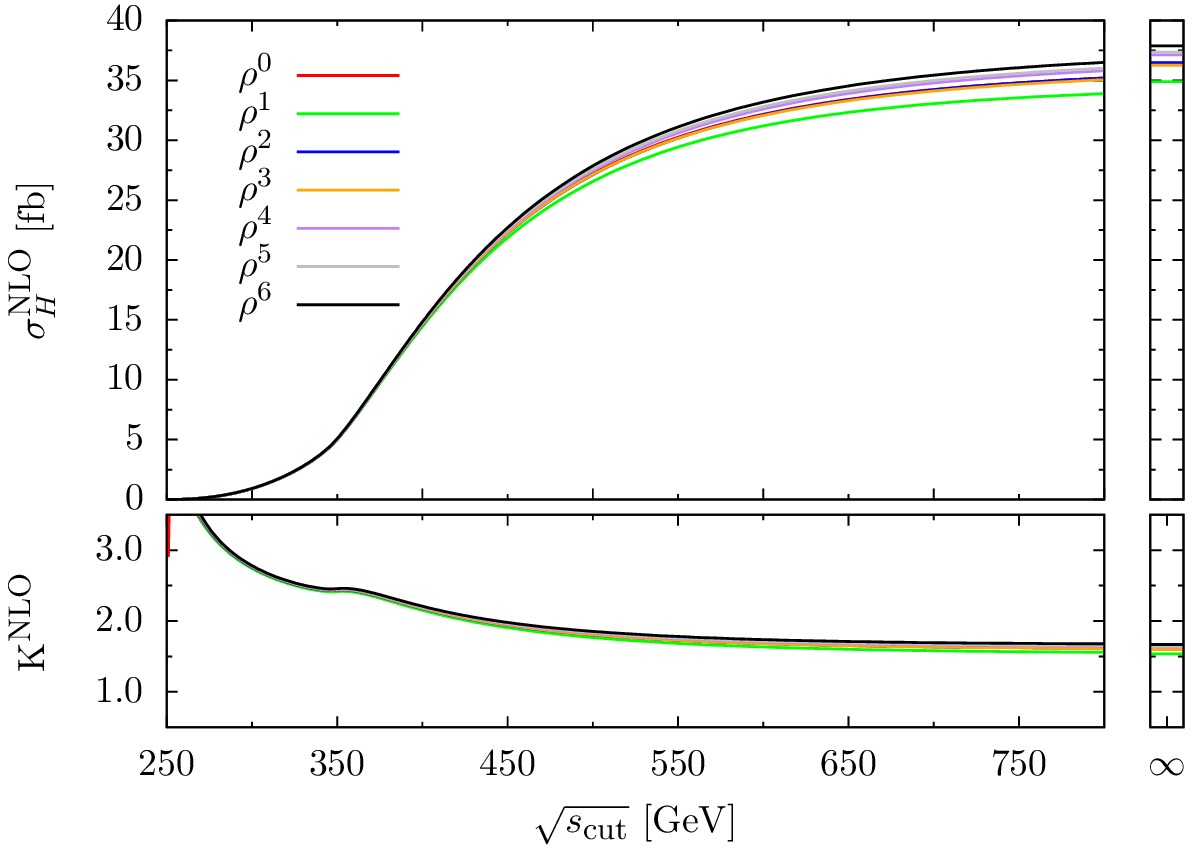}
    \\
    \caption{\label{fig::sig_hadr}
      NLO hadronic cross section and $K$ factor as a function of $\sqrt{s_{\rm cut}}$.}
  \end{center}
\end{figure}

Fig.~\ref{fig::sig_hadr} shows the hadronic cross section $\sigma_H$ for Higgs
boson pair production including NLO corrections as a function of $\sqrt{s_{\rm
    cut}}$ which is a technical upper cut on the partonic center-of-mass
collision energy. It is introduced via
\begin{eqnarray}
  \sigma_H(s_H, s_{\rm cut}) &=& \int_{4m_H^2/s_H}^1
  {\rm d} \tau \, \left(\frac{ {\rm d} {\cal L}_{gg} }{ {\rm d} \tau }\right)(\tau) \,
  \sigma(\tau s_H) \, 
  \theta(s_{\rm cut} - \tau s_H)
  \,,
\end{eqnarray}
where the luminosity function is given by
\begin{eqnarray}
  \left(\frac{ {\rm d} {\cal L}_{gg} }{ {\rm d} \tau }\right)(\tau) &=&
  \int_0^1 {\rm d} x_1 \int_0^1 {\rm d} x_2
  f_g(x_1) f_g(x_2) \delta(\tau - x_1 x_2)
  \,.
\end{eqnarray}
$f_g(x)$ are the gluon distribution functions in the $\overline{\rm MS}$
scheme.  Note that in the soft limit $\sqrt{s_{\rm cut}}$ is a good
approximation to $Q^2$.  The various lines in Fig.~\ref{fig::sig_hadr}
correspond to the inclusion of different orders in $\rho$ at NLO.  For
convenience we show on the right end of Fig.~\ref{fig::sig_hadr} the total
cross section for $\sqrt{s_H}=14$~TeV.  Note that the approximation used for the
computation of the $\rho^n$ terms is not valid for large values of
$\sqrt{s}_{\rm cut}$ (neither is the effective-theory result). However, it can
be used as an estimate of the mass correction terms.  Using the spread as an
estimate for the uncertainty we conclude that a finite top mass induces a
$\pm10$\% uncertainty on top of the infinite top quark mass result.

The lower panel of Fig.~\ref{fig::sig_hadr} shows the hadronic $K$ factor
which is obtaind from Eq.~(\ref{eq::K_NLO}) by replacing $\sigma$ by
$\sigma_H$ and using NLO PDFs in the numerator and LO PDFs in the
denominator. $K^{\rm NLO}$ raises close to threshold, however, for
$\sqrt{s_{\rm cut}}\gsim 500$~GeV one observes a flat behaviour of $K^{\rm
  NLO}\approx 1.6$ (for $\mu=2m_H$).

\begin{figure}[t]
  \begin{center}
    \includegraphics[width=1.0\columnwidth]{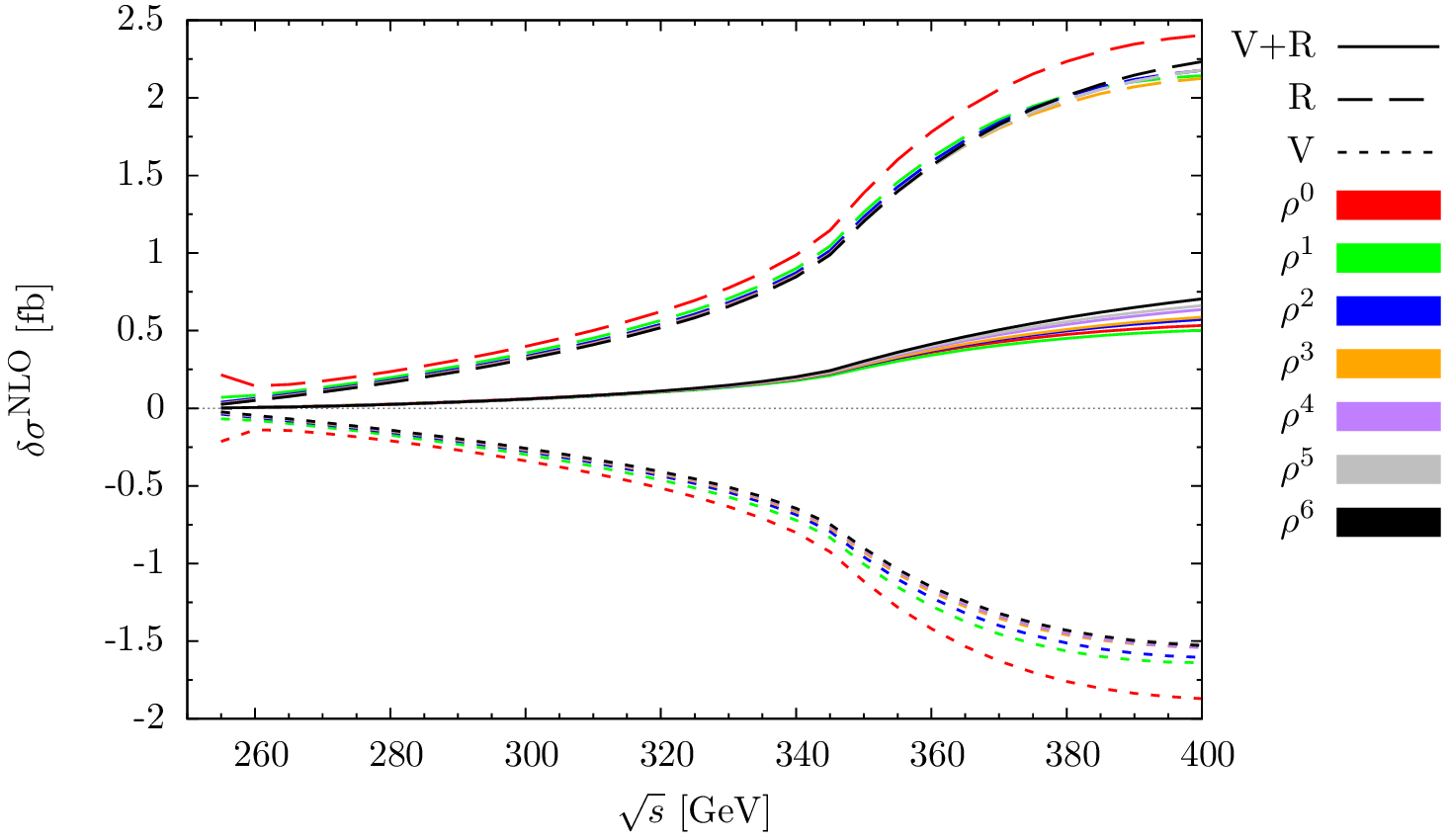}
    \\
    \caption{\label{fig::sig_part_real_virt}
      Splitting of partonic cross section (solid lines) into real (uppser
      dashed lines) and virtual (lower dotted lines) contributions
      (see text for details).}
  \end{center}
\end{figure}

Top quark mass effects to double Higgs boson production have also
been considered in Ref.~\cite{Maltoni:2014eza}. In the approximation used
in that reference the real corrections are treated exactly, however, the
infinite top quark mass approximation is used for the virtual corrections.
A decrease of the cross section by about 10\% due to finite top quark mass is
reported.

In Fig.~\ref{fig::sig_part_real_virt} we split\footnote{Note the the
  individual terms contain $1/\epsilon$ poles which cancel in the sum.  In
  Fig.~\ref{fig::sig_part_real_virt} the finite contributions are plotted.}
the partonic results for the solid lines of Fig.~\ref{fig::sig_part} (which
corresponds to the factorization according to Eq.~(\ref{eq::nlo_fact1})) into
virtual corrections (including the ultra-violet counterterms; dotted lines) and the
real-radiation part which include the contributions from the splitting
functions (dashed lines).

We observe that for $\sqrt{s}\lsim 400$~GeV, the region where our approximation
is valid, the top quark mass corrections to the real radiation part (upper
dashed curves) reduce the infinite top result by about 10\%, in agreement with
the observations of Ref.~\cite{Maltoni:2014eza}.  On the other hand, the top
mass effects to the virtual contribution leads to a positive shift as compared
to the effective theory result.  Summing real and virtual corrections leads to
an overall positive effect from top mass corrections as can be seen by the
solid curves, see also Fig.~\ref{fig::sig_part_2}.  Note that up to $\sqrt{s}
\approx 400$~GeV top mass corrections are dominated by the $1/M_t^2$ terms.

%- }}}
%- {{{ Top quark mass corrections at NNLO:

\section{\label{sec::nnlo}Top quark mass corrections at NNLO}

In this Section we compute the virtual corrections to the NNLO cross
section including top quark mass corrections.  Afterwards we construct
$\Sigma_{\rm fin}$ as described in Section~\ref{sec::fac} and use
Eq.~(\ref{eq::G_SV}) to evaluate the partonic and hadronic cross
sections including $1/M_t^4$ correction terms. 

\subsection{Calculation}

We have applied two methods to compute the virtual corrections. In the first
one we consider the amplitudes for $gg\to HH$ up to three-loop order and in
the second one the forward scattering amplitude is considered, which, after
taking the imaginary part, directly leads to the total cross section. In both
cases we perform an asymptotic expansion for large top quark mass. The first
approach has the advantage that it is straightforward to introduce Higgs boson
decays whereas the second approach can immediately be applied to real
corrections.

\subsubsection{Amplitude $gg\to HH$}

\begin{figure}[t]
  \begin{center}
    \centering
    \raisebox{-0.5\height}{\includegraphics{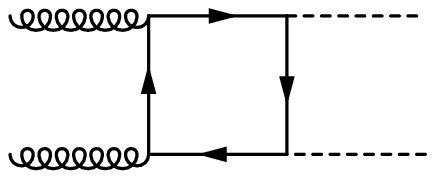}}\hspace{1cm}
    \raisebox{-0.5\height}{\includegraphics{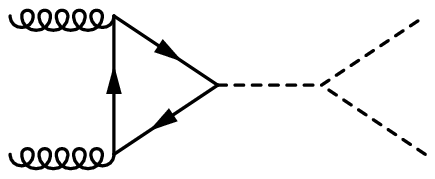}}\\[10pt]
    \raisebox{-0.5\height}{\includegraphics{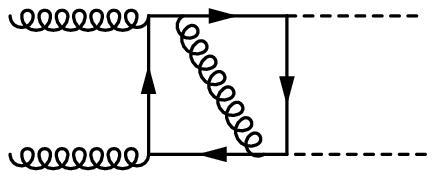}}\hfill
    \raisebox{-0.5\height}{\includegraphics{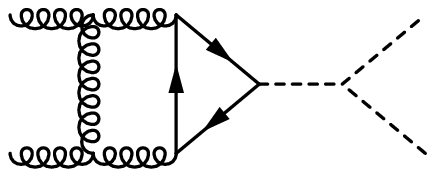}}\hfill
    \raisebox{-0.5\height}{\includegraphics{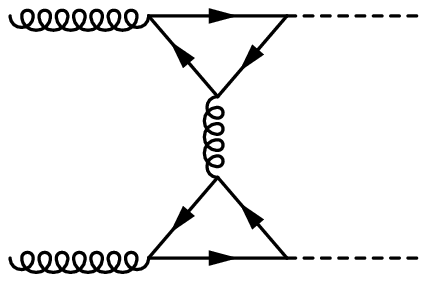}}\\[10pt]
    \raisebox{-0.5\height}{\includegraphics{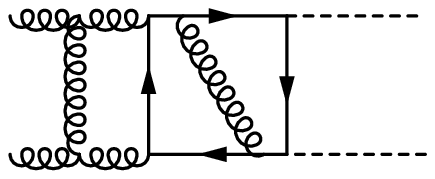}}\hfill
    \raisebox{-0.5\height}{\includegraphics{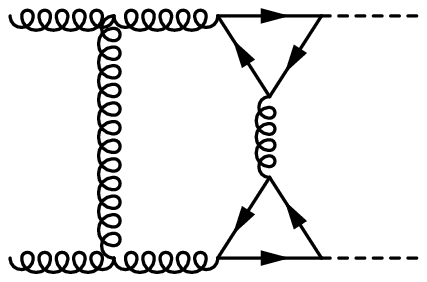}}\hfill
    \raisebox{-0.5\height}{\includegraphics{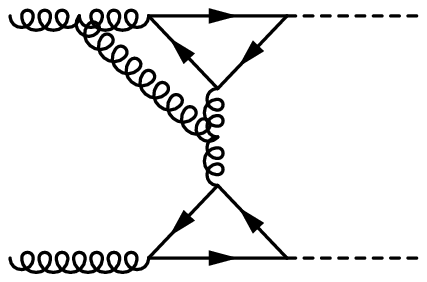}}\\[10pt]
    \caption{\label{fig::gghh}
      One-, two- and three-loop 
      Feynman diagrams contributing to the process $gg\to HH$.
      Solid lines refer to top quarks, curly lines to        
      gluons and dashed lines to Higgs bosons.}
  \end{center}
\end{figure}

\begin{figure}[t]
  \begin{center}
      \includegraphics[scale=1]{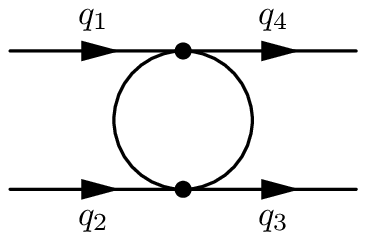}\qquad
      \includegraphics[scale=1]{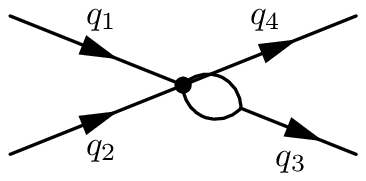}\qquad
      \includegraphics[scale=1]{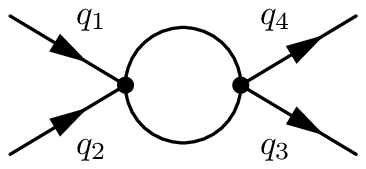}\\
      \includegraphics[scale=1]{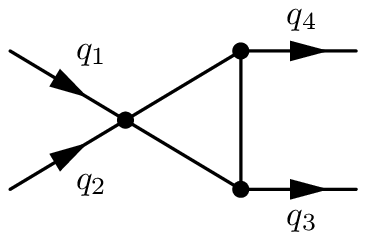}\qquad
      \includegraphics[scale=1]{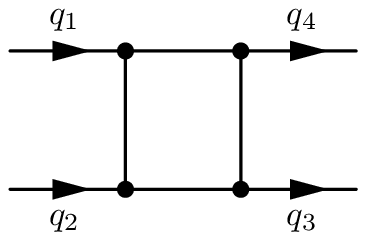}\qquad
      \includegraphics[scale=1]{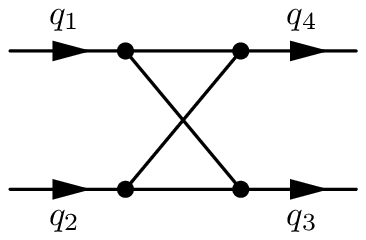}\\
      \includegraphics[scale=1]{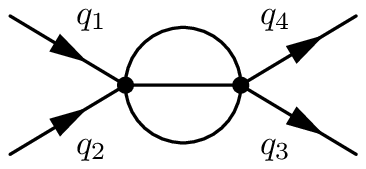}\qquad
      \includegraphics[scale=1]{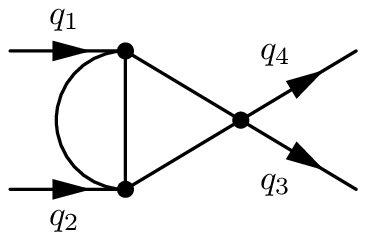}\qquad
      \includegraphics[scale=1]{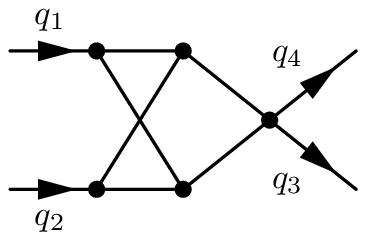}\\
      \includegraphics[scale=1]{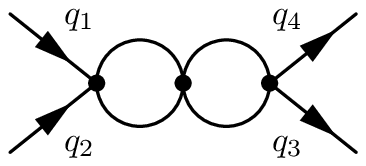}\qquad
      \\
    \caption{\label{fig::gghh_ml}
      One- and two-loop 
      master integrals needed after applying asymptotic expansion to
      the amplitude $gg\to HH$. All internal lines are massless,
      $q_1^2=q_2^2=0$, and $q_3^2=q_4^4=m_H^2$.}
  \end{center}
\end{figure}

NNLO calculations require corrections up to three-loop order to the process
$gg\to HH$.  Typical contributing Feynman diagrams at LO, NLO and NNLO are
shown in Fig.~\ref{fig::gghh}. They are generated with the help of {\tt
  qgraf}~\cite{Nogueira:1991ex}.  Note that in this approach no contributions
with external ghosts have to be considered since we project on physical
states.  The transformation to {\tt FORM}~\cite{Kuipers:2012rf} code is done
with the program {\tt q2e}~\cite{Harlander:1997zb,Seidensticker:1999bb} and
the asymptotic expansion for large top quark masses is realized with the help
of {\tt exp}~\cite{Harlander:1997zb,Seidensticker:1999bb}. After expanding the
identified hard subgraphs in the small quantities one arrives at one-scale vacuum
integrals up to three loops and massless one- and two-loop four-point diagrams
with two massless and two massive external momenta.  The vacuum integrals are
computed using {\tt MATAD}~\cite{Steinhauser:2000ry}. In the case of the four-point
integrals we apply {\tt FIRE}~\cite{Smirnov:2013dia,Smirnov:2014hma} to
express them as linear combinations of master integrals. The latter are shown
in Fig.~\ref{fig::gghh_ml} where we have $q_1^2=q_2^2=0$, and
$q_3^2=q_4^4=m_H^2$. Analytic results for all master integrals can be
found in the literature (see, e.g.,
Refs.~\cite{Birthwright:2004kk,Ellis:2007qk,Chavez:2012kn}). In this
paper we only show results for the
triangle graph in the second line of Fig.~\ref{fig::gghh_ml} since for our purpose
the representations given in
Refs.~\cite{Birthwright:2004kk,Chavez:2012kn} are less suited. We use instead
\begin{align*}
  I_1(4) &= \frac{1}{s \sqrt{\delta}} 
  \left(\frac{\mu^2}{m_H^2}\right)^\epsilon
  \Bigg\{G_o(-1; x) G_o(0; y) -
    G_o(0; y) G_o(-1/y; x) + 2 G_o(-1, 0; x) \\
& - 2 G_o(-1/y, 0; x) +\epsilon \Bigg[-2i\pi G_o(-1, 0; x) - G_o(-1/y; x)G_o(0, 0; y)\\
& + G_o(-1; x) \Big(-i\pi G_o(0; y) + G_o(0, 0; y)\Big) + 2i \pi G_o(-1/y, 0;x)\\
& + G_o(0; y) \Big(i \pi G_o(-1/y; x) + G_o(-1, -1; x) + 2G_o(-1, 0; x) \\
& - G_o(-1, -1/y; x)  + G_o(-1/y, -1; x) - 2G_o(-1/y, 0; x) -
          G_o(-1/y, -1/y; x)\Big)\\
&  + 2 G_o(-1, -1, 0; x) + 4 G_o(-1, 0, 0; x) -
        2G_o(-1, -1/y, 0; x) + 2 G_o(-1/y, -1, 0; x) \\
&- 4G_o(-1/y, 0, 0; x) - 2G_o(-1/y, -1/y, 0; x) \Bigg] + \mathcal{O}(\epsilon^2) \Bigg\}\,,
\numberthis
\label{eq::I14}
\end{align*}
with
\begin{align}
  y &= \frac{1}{x^2} \ , & x & = \frac{1+\sqrt{\delta}}{1 -
    \sqrt{\delta}} \ , &  \delta &= 1 - \frac{4 m_H^2}{s} \,.
\end{align}
$G_o(\{w_i\};z)$ are Goncharov Polylogarithms~\cite{Goncharov:1998kja} with weight
$\{w_i\}$ and argument $z$ defined through
\begin{align}
  G_o(w_1, w_2, \dots, w_n; x) &=  \int_0^x \mathrm{d} t \frac{1}{t - w_1} G_o (w_2, \dots, w_n; t) \,,
\end{align}
with $w_i, x \in \mathbb{C}$ and
\begin{align}
  G_o(\vec{0}_n;x) &= \frac{1}{n!} \log^n x\,.
% & \vec{0}_n &= (\underbrace{0, \dots, 0}_{n \textrm{ times}})\,. 
\end{align}

The functions of the $\epsilon^0$ term in Eq.~(\ref{eq::I14})
can be expressed in terms of logarithms and dilogarithms via
\begin{align*}
G_o(0; y) & = \log(y) \ , \\
G_o(-1; x) & = \log(1+x) \ , \\
G_o(-1/y; x) & = \log (1+xy) \ , \\
G_o(-1, 0; x) &= \log(x) \log(1+x) + \mathrm{Li}_2(-x) \ , \\
G_o(-1/y, 0; x) &= \log(x) \log(1+xy) + \mathrm{Li}_2(-xy) \, .
\numberthis
\end{align*}
We have cross checked the numerical result for the $\epsilon^0$ and $\epsilon^1$ terms of
$I_1(4)$ in Eq.~(\ref{eq::I14})
against {\tt FIESTA}~\cite{Smirnov:2013eza}.

Using this method we have computed terms up
to order $1/M_t^{12}$ at NLO~\cite{Grigo:2013rya,Grigo:2013xya,Grigo:2014oqa}
and terms up to order $1/M_t^4$ at NNLO.  As an important check we have
computed the $1/M_t^2$ corrections for general QCD gauge parameter which drops
out in the final expression.

From the calculation of $gg\to HH$ one obtains in a first step 
results for ${\rm d}\sigma/{\rm d}t$.
Integration over phase space then leads to ${\rm d}\sigma/{\rm d}Q^2$.
For the results in Section~\ref{sec::impr_nnlo} this integration is performed
numerically.

\subsubsection{Amplitude $gg\to gg$}

\begin{figure}[t]
  \begin{center}
    \centering
    \includegraphics{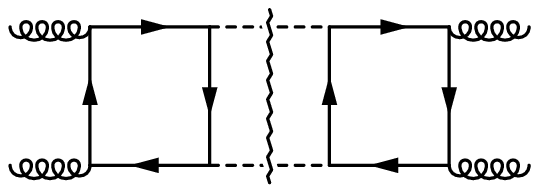}\hspace{1cm}
    \includegraphics{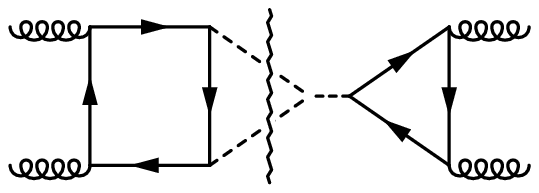}\\[10pt]
    \includegraphics{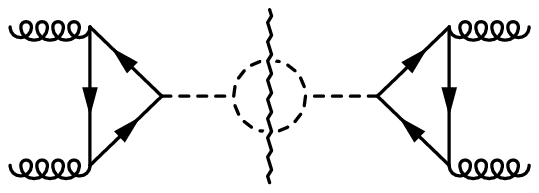} \\[10pt]
    \includegraphics{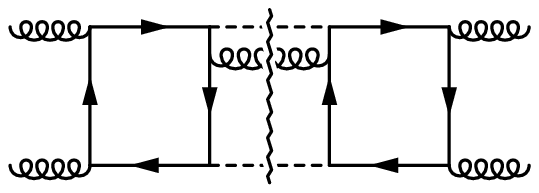}\hspace{1cm}
    \includegraphics{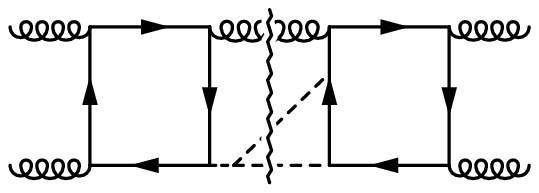}\\[10pt]
    \includegraphics{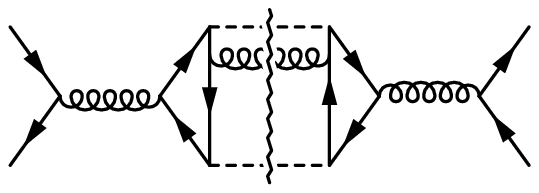}\hspace{1cm}
    \includegraphics{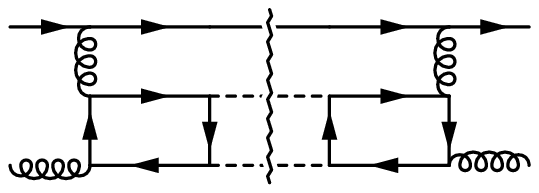} \\[10pt]
    \includegraphics{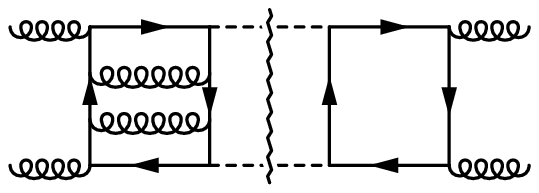}\hspace{1cm}
    \includegraphics{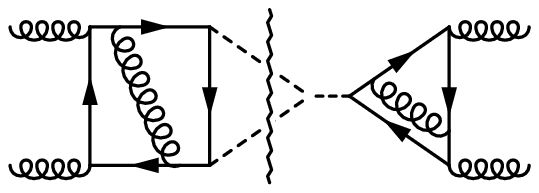} \\[10pt]
    \caption{\label{fig::gg2gg}
      LO, NLO and NNLO Feynman diagrams needed for the 
      forward scattering amplitude $gg\to gg$.
      Solid lines refer to top quarks, curly lines to        
      gluons and dashed lines to Higgs bosons.
      At NNLO only virtual contributions are shown.
      Wavy lines denote the cuts.}
  \end{center}
\end{figure}

The second method is based on the use of the optical theorem
in analogy to the NLO calculation performed in Ref.~\cite{Grigo:2013rya}.
This method serves as an important cross check.  In the following we
provide some of the technical details
\begin{itemize}
\item The amplitudes for $gg\to gg$ are generated
  with the help of {\tt qgraf}~\cite{Nogueira:1991ex}.
\item In a first step about 17 million diagrams are generated.  However, most of
  them do not contain a cut through exactly two Higgs bosons. For this reason
  we post-process the {\tt qgraf} output and
  filter~\cite{Grigo:2014oqa,diss_Hoff} the amplitudes describing the virtual
  corrections to $gg\to HH$.  Typical Feynman diagrams are shown in
  Fig.~\ref{fig::gg2gg}.
\item
  {\tt FORM}~\cite{Vermaseren:2000nd,Kuipers:2012rf} code is then generated by
  passing the output via {\tt
    q2e}~\cite{Harlander:1997zb,Seidensticker:1999bb}, which transforms
  Feynman diagrams into Feynman amplitudes, to {\tt
    exp}~\cite{Harlander:1997zb,Seidensticker:1999bb}.
\item 
  Our in-house {\tt FORM} code applies projectors $(-g_{\mu\nu})$
  for each pair of external gluons which includes also non-physical
  degrees of freedom in the sum. Thus also contributions with ghosts in the
  initial state have to be considered.
  Note that this is in contrast to single Higgs boson production which 
  has no virtual contributions with ghosts in the initial state.
\end{itemize}

\begin{figure}[t]
  \begin{center}
    \includegraphics[width=0.8\columnwidth]{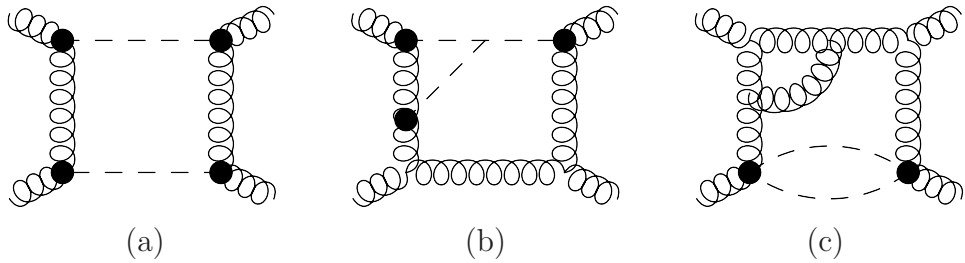}
    \\
    (a) \hspace*{9em} (b) \hspace*{9em} (c)
    \caption{\label{fig::gg2gg_ae}
      Resulting Feynman diagrams after shrinking the top quark loops
      to a point according to the rule of asymptotic expansion.
      The blob represents one-loop vacuum integrals.}
  \end{center}
\end{figure}

\begin{figure}[t]
  \begin{center}
    \includegraphics[scale=0.4]{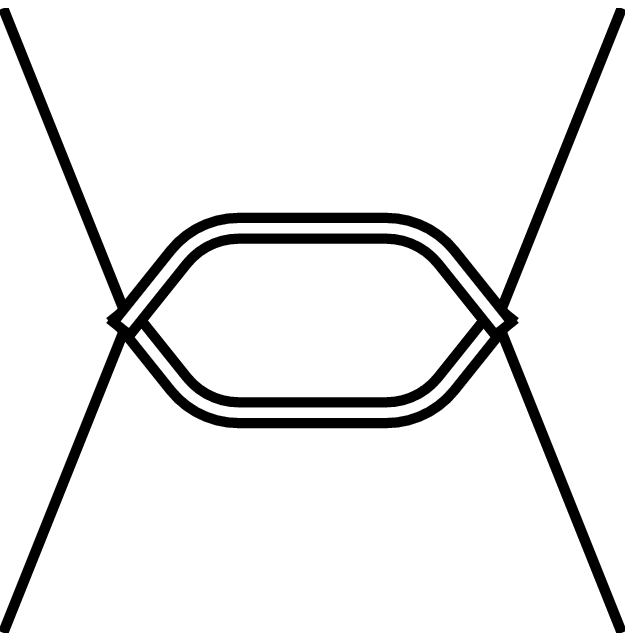}
    \includegraphics[scale=0.4]{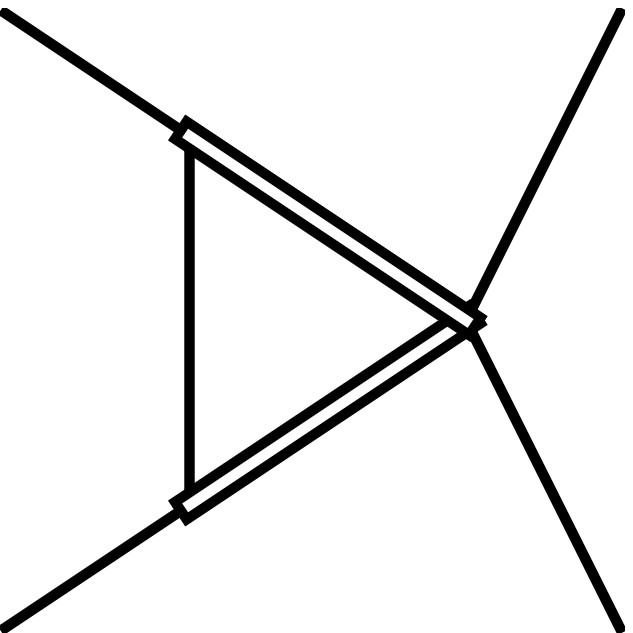}
    \includegraphics[scale=0.4]{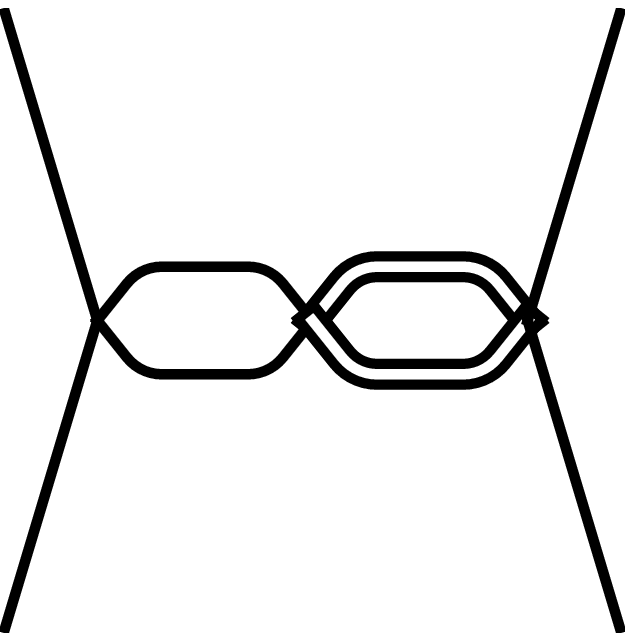}
    \includegraphics[scale=0.4]{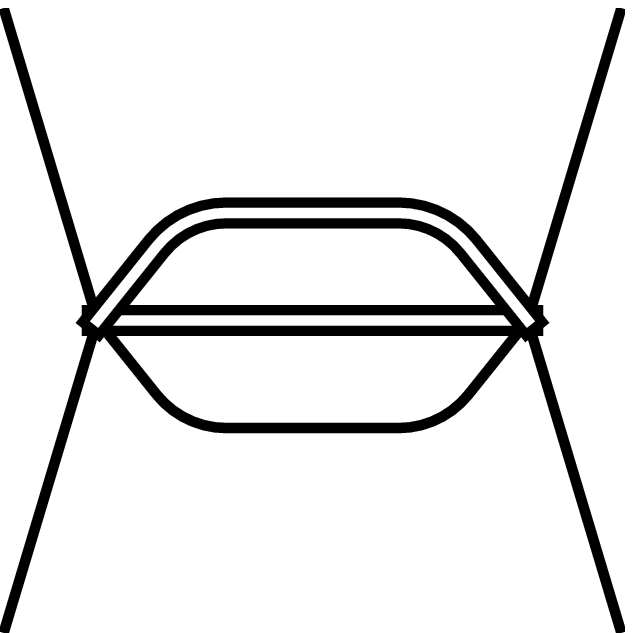}
    \includegraphics[scale=0.4]{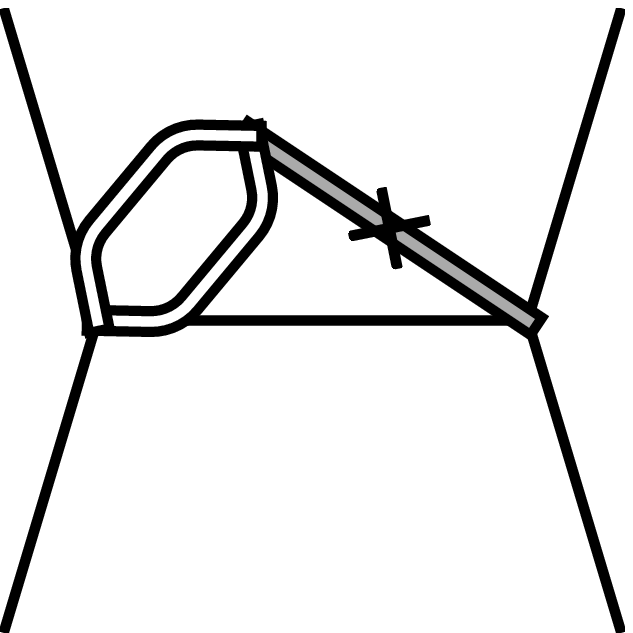}
    \includegraphics[scale=0.4]{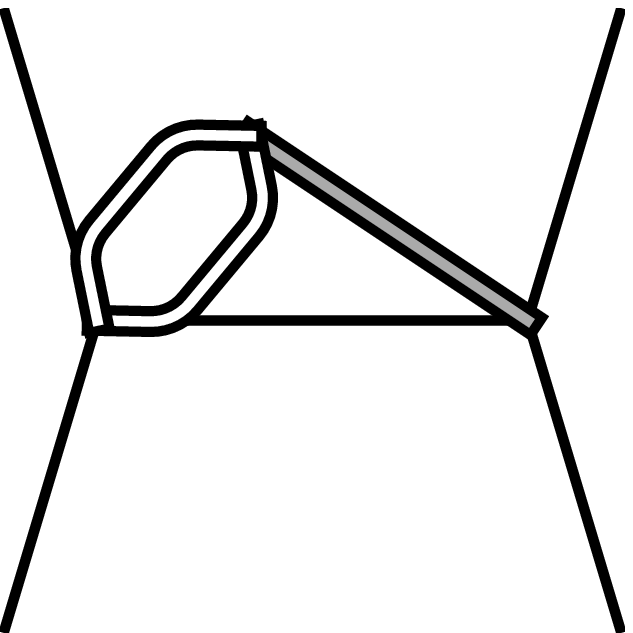}\\
    \includegraphics[scale=0.4]{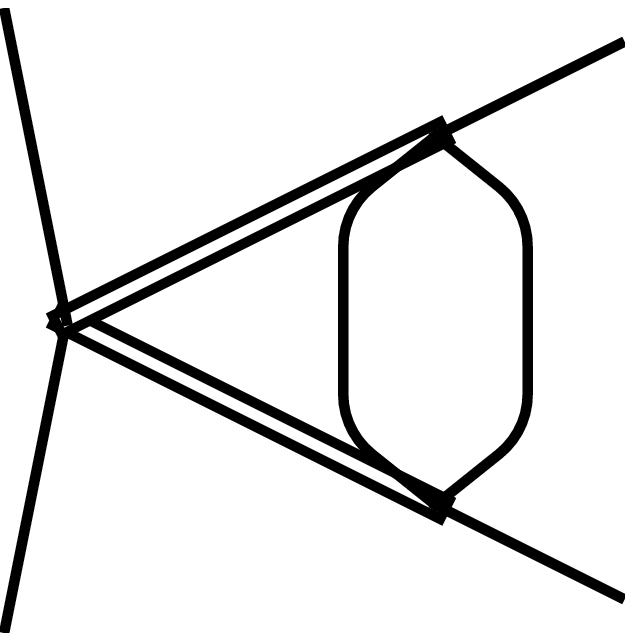}
    \includegraphics[scale=0.4]{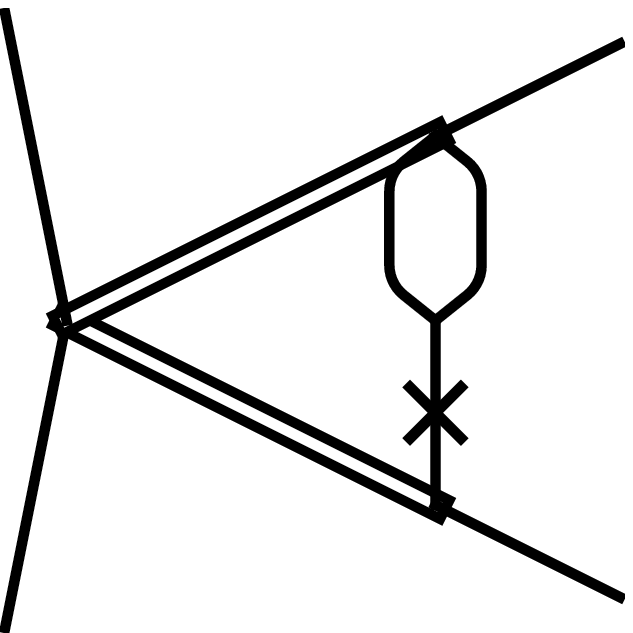}
    \includegraphics[scale=0.4]{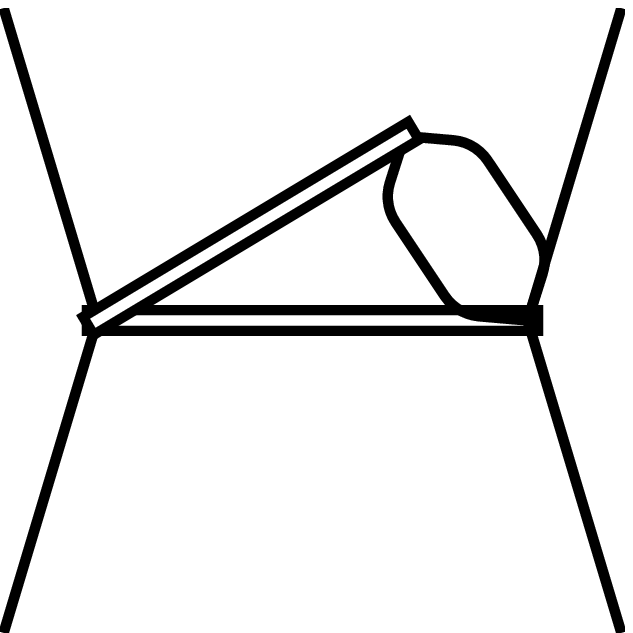}
    \includegraphics[scale=0.4]{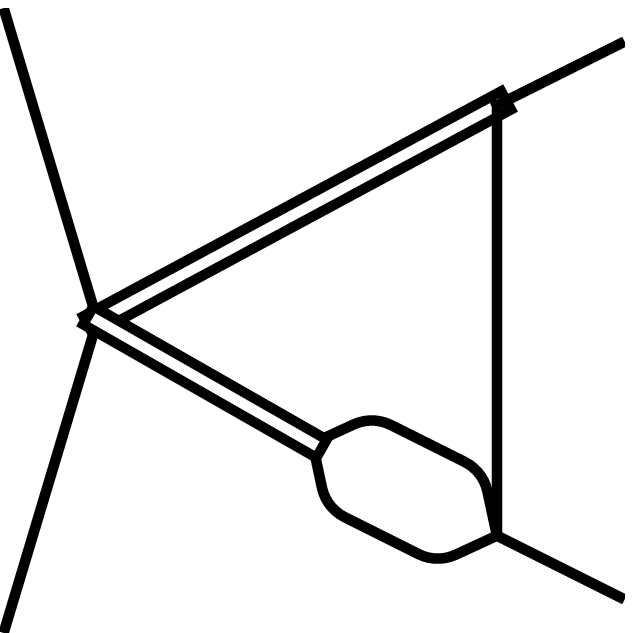}
    \includegraphics[scale=0.4]{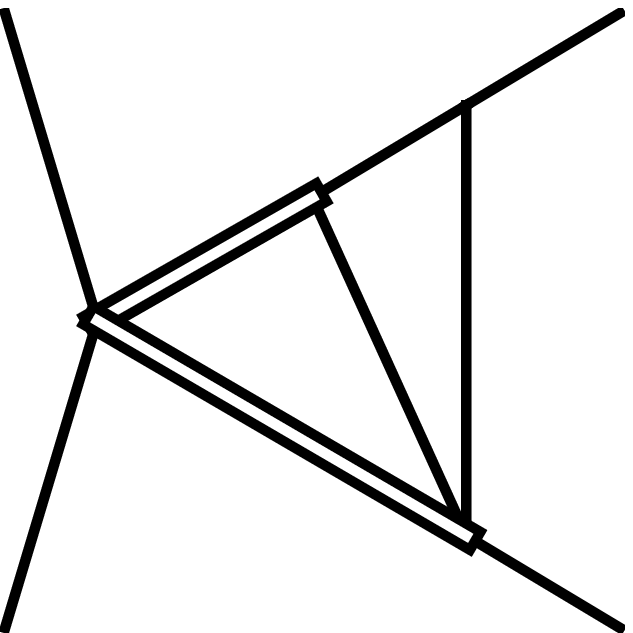}
    \includegraphics[scale=0.4]{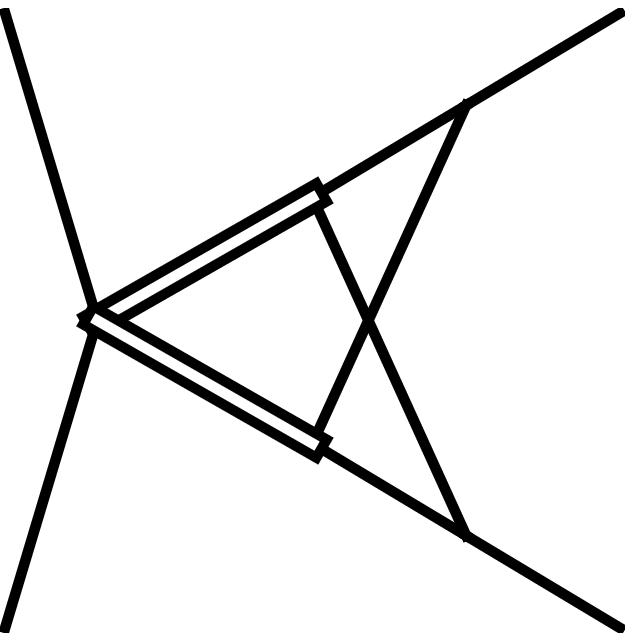}\\
    \includegraphics[scale=0.4]{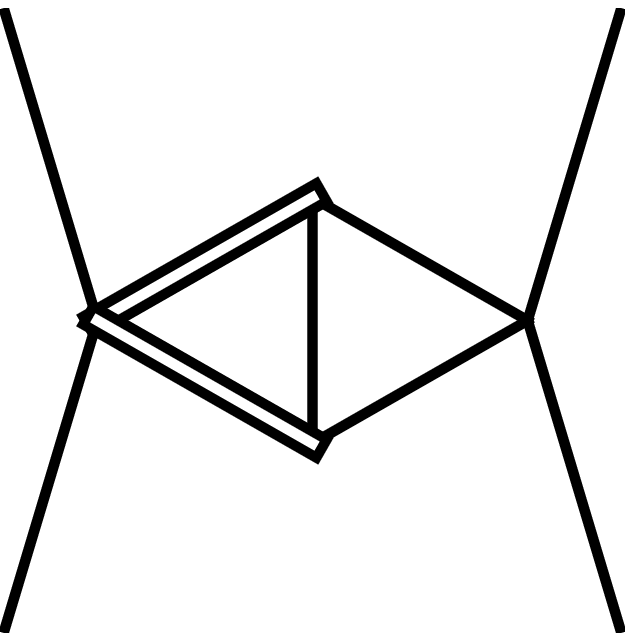}
    \includegraphics[scale=0.4]{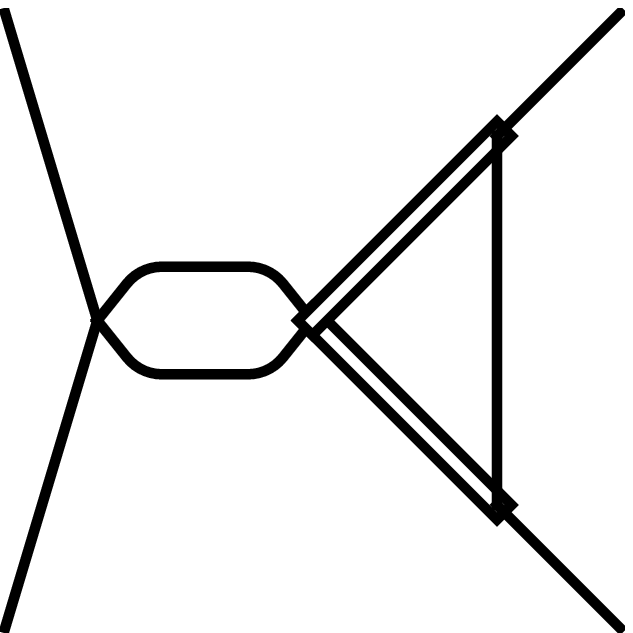}
    \includegraphics[scale=0.4]{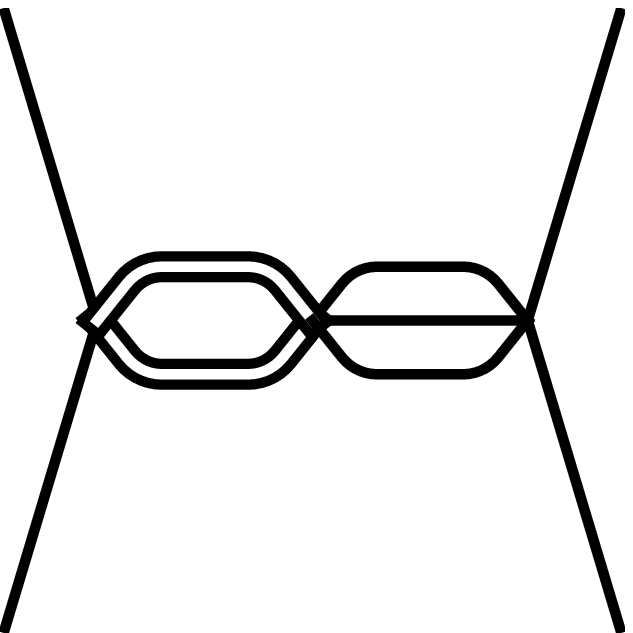}
    \includegraphics[scale=0.4]{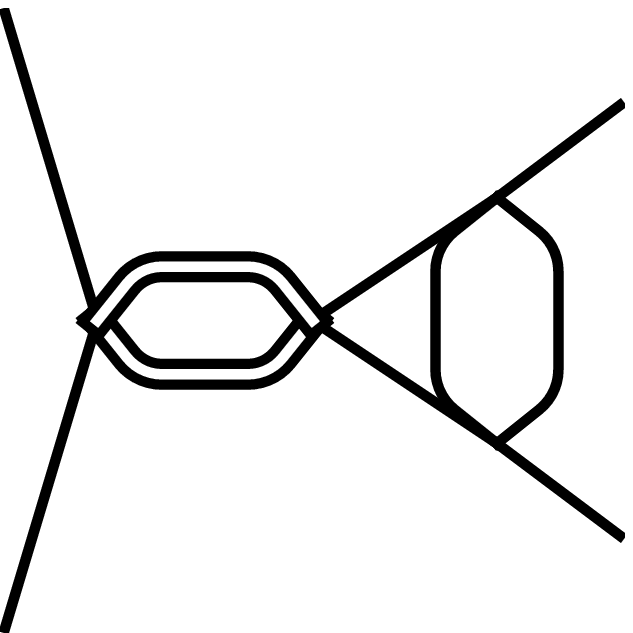}
    \includegraphics[scale=0.4]{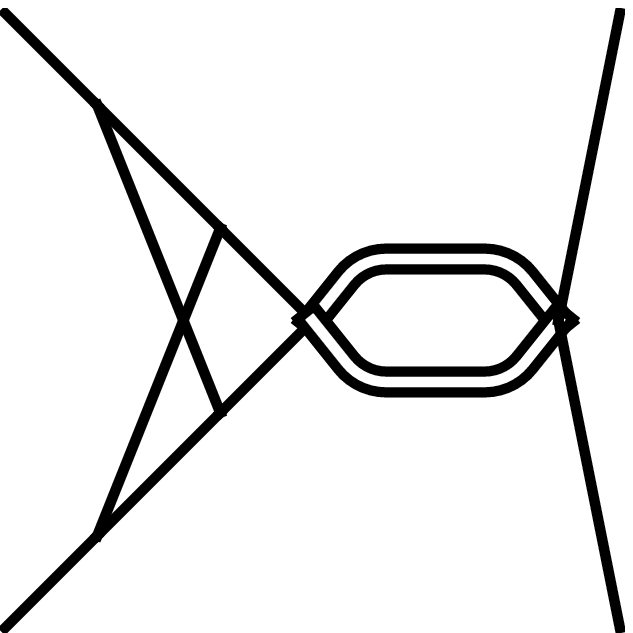}
    \includegraphics[scale=0.4]{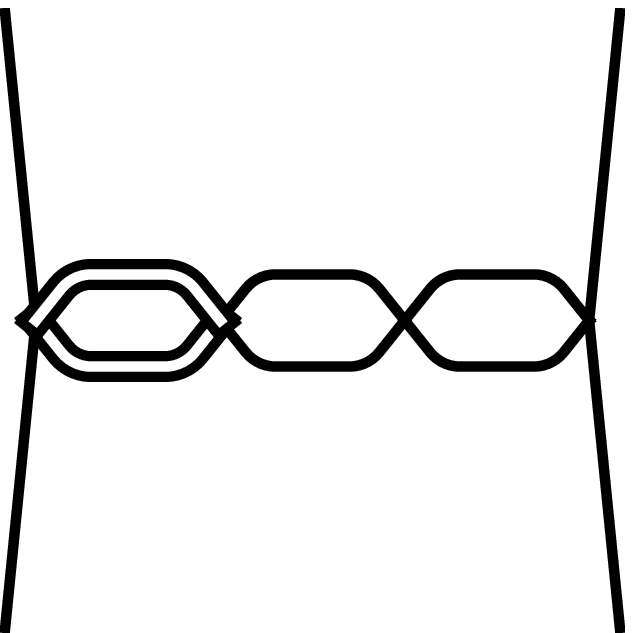}
    \caption{\label{fig::gg2gg_MI} Phase space master integrals occurring in
      the amplitude for $gg\to gg$.  The first line contains LO and NLO
      integrals. The integrals in line two and three are needed for the NNLO
      virtual corrections. Single and double lines represent massless and
      Higgs propagators, respectively. Double lines with gray-shaded
      interspace (last two diagrams in first row)
      correspond to Higgs boson propagators which shall not be
      cut. A cross marks an inverse propagator.}
  \end{center}
\end{figure}

The application of the asymptotic expansion for large top quark mass
leads to a factorization of the five-loop integrals into the following
contributions:
\begin{enumerate}
\item four one-loop vacuum integrals times one-loop integrals which are
  already present at LO, see Fig.~\ref{fig::gg2gg_ae}(a);
\item three one-loop vacuum integrals times two-loop integrals,
  see Fig.~\ref{fig::gg2gg_ae}(b);
\item two one-loop vacuum integrals times three-loop integrals,
  see Fig.~\ref{fig::gg2gg_ae}(c);
\item two-loop times one-loop vacuum integrals times two-loop integrals;
\item three-loop times one-loop vacuum integrals times one-loop integrals.
\end{enumerate}
The mass scale in the vacuum integrals is given by the top quark.  They are
again computed with the help of {\tt MATAD}~\cite{Steinhauser:2000ry}.  For
the remaining integrals, which depend on $\delta$, we use 
the in-house programs {\tt rows}~\cite{diss_Hoff} and {\tt
  TopoID}~\cite{Grigo:2014oqa,diss_Hoff} to perform the reduction to 
master integrals. The latter are depicted in Fig.~\ref{fig::gg2gg_MI}.

All three-loop master integrals factorize into a two-loop form factor
contribution and the one-loop master integral of the LO calculation.
From the latter only the imaginary part is needed which is well known.
The results for the two-loop form factor integrals
can, e.g., be found in Refs.~\cite{Gehrmann:2005pd}.

The two-loop master integrals are more involved.  A numerical calculation
would probably be possible, however, we follow the approach outlined in
Ref.~\cite{Grigo:2013rya} for the NLO master integrals and perform an
expansion in $\delta=1-4m_H^2/s$ [cf. Eq.~(\ref{eq::delta})].  
All integrals contain a massless sub-loop for which
analytic expressions are known. The massless two-point function can be
expressed in terms of $\Gamma$ functions and results for the triangle with two
massless external legs and the (crossed) box can be found in
Ref.~\cite{Ellis:2007qk}. 
Analytic expressions for the triangle diagram with squared external momenta
$s, m_H^2, m_H^2$ are given in Eq.~(\ref{eq::I14}).  After expanding in
$\delta$ the remaining phase-space integration can be performed
analytically. We have computed expansion terms up to order $\delta^{10}$
and found agreement with the results obtained in the previous subsection which
for this purpose have also been integrated analytically after performing the
expansion in $\delta$.

The approach based on the optical theorem requires special care
in the treatment of the imaginary parts originating from the 
two-loop form factor diagrams. Such contributions either
correspond to $|{\cal M}^{\rm NLO}|^2$ or 
$({\cal M}^{\rm LO} {\cal M}^{\star \rm NNLO} + 
{\cal M}^{\rm NNLO} {\cal M}^{\star \rm LO})$.
In the former case the two-loop integrals originate from the product of 
two one-loop contributions containing factors
$(-1+i0)^\epsilon$ and $(-1-i0)^\epsilon$, respectively,
which finally leads to $(-1+i0)^\epsilon (-1-i0)^\epsilon=1$.
In the other case one has
$(-1+i0)^{2\epsilon} + (-1-i0)^{2\epsilon} = 
1-4\pi^2\epsilon^2 + {\cal O}(\epsilon^3)$.
The corresponding discussion for single Higgs production can be found 
in Ref.~\cite{Pak:2011hs}.

Note that the approach based on the $gg\to HH$ amplitudes 
leads to simpler intermediate expressions. Thus, it is
possible to allow for a general QCD gauge parameter $\xi$ when computing the
$1/M_t^2$ terms. Furthermore, also the $1/M_t^4$ corrections could be
evaluated (for $\xi=0$) whereas in the optical theorem approach only
$1/M_t^2$ terms could be computed in Feynman gauge. 
However, let us mention that this approach can be used in 
a straightforward way to compute NNLO top quark mass effects to the hard
contribution of the total cross section whereas in the approach 
of the previous subsection this is less obvious.

\bigskip

To conclude this Section let us summarize our procedure to obtain the SV
corrections at NNLO. We compute virtual corrections to $gg\to HH$ including 
$1/M_t^4$ corrections. Note that we have 
${\rm d}\sigma/{\rm d}Q^2|_{\rm virt} \sim \delta(Q^2-s)$. 
Using Eq.~(\ref{eq::sig_fin}) we construct 
$\sigma_{\rm fin}$ which enters $G_{\rm SV}$ in
Eq.~(\ref{eq::G_SV}). The differential and total cross section is then
obtained with the help of Eqs.~(\ref{eq::G}) and~(\ref{eq::sigma_tot}).

%- }}}
%- {{{ Improving NNLO:

\section{\label{sec::impr_nnlo}Improving NNLO}

In this Section we discuss the effect of the $1/M_t^2$ and $1/M_t^4$ terms on
the NNLO cross section for the production of Higgs boson pairs. The cross
section in the infinite top quark mass limit has been computed in
Ref.~\cite{deFlorian:2013uza}. In Ref.~\cite{Grigo:2014jma} the three-loop
matching coefficient has been added, completing the NNLO prediction, 
and the virtual corrections from
Ref.~\cite{deFlorian:2013uza} have been cross checked.

\begin{figure}[t]
  \begin{center}
    \includegraphics{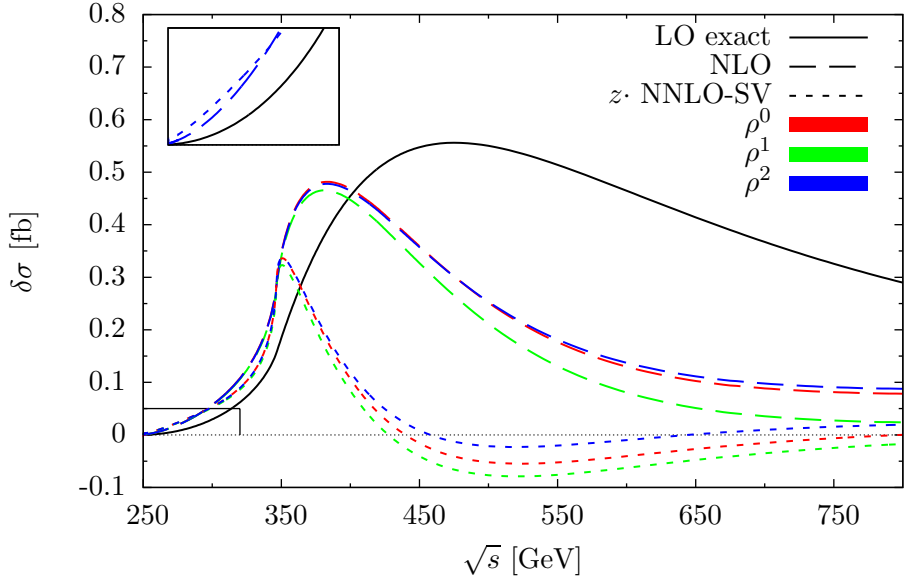}
    \\
    \caption{\label{fig::sig_part_lo_nlo_nnlo}
      Comparison of the LO, NLO and NNLO contributions to the partonic cross
      section. At LO the exact result is shown and at NLO and NNLO the first
      three terms in the large-$M_t$ expansion are shown. For all curves the 
      NNLO-value for $\alpha_s$ is used. For the renormalization and
      factorization scale we use $\mu=2m_H$.
    }
  \end{center}
\end{figure}

The results for the virtual corrections computed in
Section~\ref{sec::nnlo} are inserted in the formalism described in
Section~\ref{sec::fac} to construct the quantity $\sigma_{\rm fin}$
which enters Eq.~(\ref{eq::G_SV}).
The result for the partonic cross section is shown in
Fig.~\ref{fig::sig_part_lo_nlo_nnlo} as a function of the partonic
center-of-mass energy where the exact LO result is
compared with NLO and NNLO. At NLO and NNLO three terms in the mass
expansion are shown.\footnote{In this plot we only include mass
  corrections up to order $\rho^2$ at NLO to have a direct comparison
  with NNLO.} Furthermore, at NNLO the SV approximation is shown for $f(z)=z$
(cf. discussion in Section~\ref{sec::nlo}). Note that the NNLO curves
peak for smaller values of $\sqrt{s}$ than at NLO and
LO. As far as the top quark mass corrections are concerned the same pattern is
observed as at NLO: the correction term of order $\rho$ decreases the infinite
top quark mass result which is overcompensated by the $\rho^2$ term 
resulting in a small positive correction.

\begin{figure}[t]
  \begin{center}
    \includegraphics{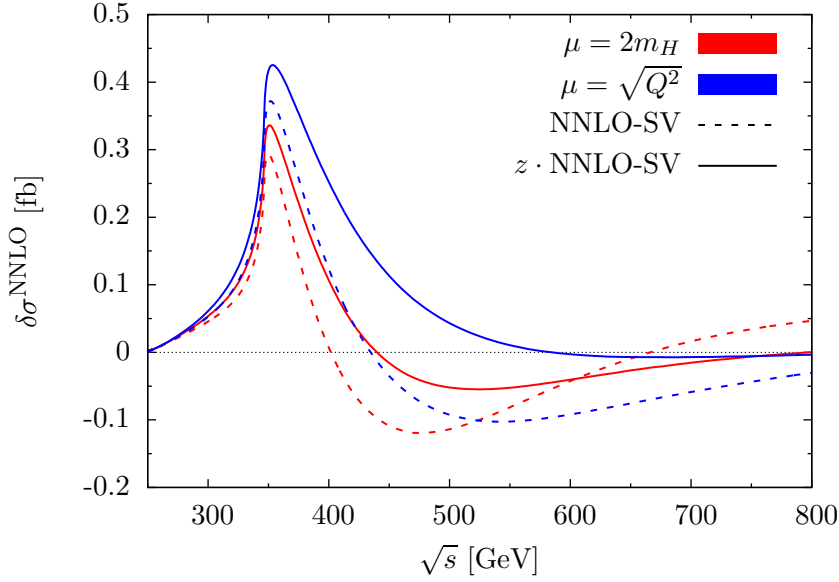}
    \\
    \caption{\label{fig::sig_part_nnlo}
      NNLO partonic cross section for different scales and 
      for $f(z)=1$ (dotted) and $f(z)=z$ (solid). 
      Only the $\rho^0$ result is shown.
    }
  \end{center}
\end{figure}

In Fig.~\ref{fig::sig_part_nnlo} we compare the NNLO-SV contribution for
two different scales, $\mu=2m_H$ and $\mu=\sqrt{Q^2}$. This plot furthermore
shows the effect of $f(z)=1$ and $f(z)=z$. Note that the choice $f(z)=z$, which we
expect to better approximate the complete result, leads to
an increase of the cross section.

\begin{figure}[t]
  \begin{center}
    \includegraphics{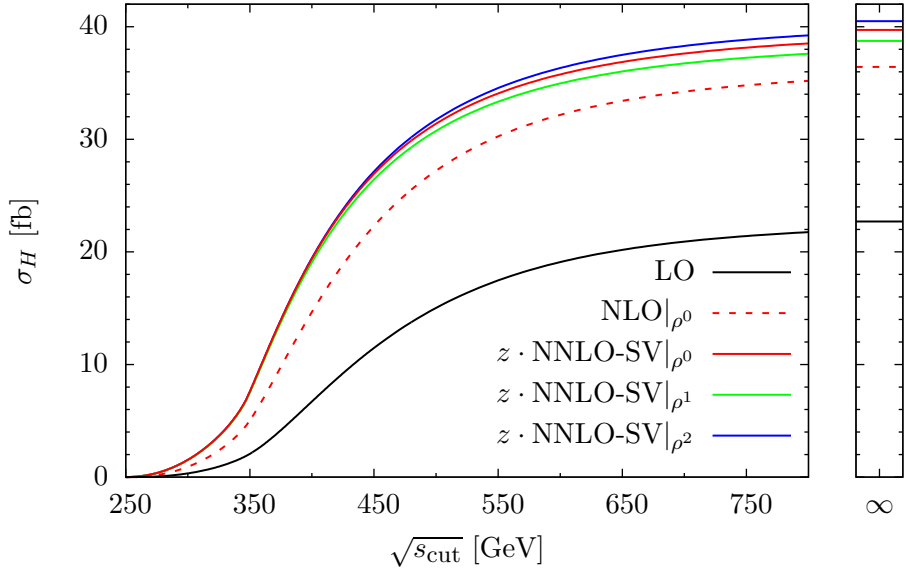}
    \\
    \caption{\label{fig::sig_hadr_lo_nlo_nnlo} Hadronic LO, NLO and NNLO-SV
      (with $f(z)=z$) cross sections as a function of $\sqrt{s_{\rm cut}}$.
      For their evaluation the respective value
      of $\alpha_s$ is used.  At LO the exact result and at NLO only the
      $\rho^0$ term is shown.  At NNLO the $\rho^0$, $\rho^1$ and $\rho^2$
      results are plotted.  The results in the right panel with ``$\infty$''
      at the bottom correspond to the prediction of the total cross section.
      For this plot $\mu=2m_H$ has been used.
    }
  \end{center}
\end{figure}

The hadronic cross section as a function of $\sqrt{s_{\rm cut}}$ is shown in
Fig.~\ref{fig::sig_hadr_lo_nlo_nnlo} for $\mu=2m_H$.  At LO the exact result
is used and the NLO curve is based on the $\rho^0$ results. Using instead the
$\rho^6$ terms leads to an upwards shift of about 5\% as can be seen in
Fig.~\ref{fig::sig_hadr}.  At NNLO three curves are shown which include terms
up to $\rho^0$, $\rho^1$ and $\rho^2$. As at NLO one observes good convergence
up to $\sqrt{s_{\rm cut}}\approx 400$~GeV. For higher values of $\sqrt{s_{\rm
    cut}}$ the $\rho^1$ curve is below and the $\rho^2$ curve above the
infinite top quark mass result leading to
a $\pm 2.5\%$ effect for the total
cross section on the rightmost part of the plot. To be conservative, we thus
estimate that the NNLO top quark mass effects lead to an uncertainty of $\pm
5\%$.  Note that the NNLO corrections amount to about 20\% of the LO result.

\begin{figure}[t]
  \begin{center}
    \includegraphics{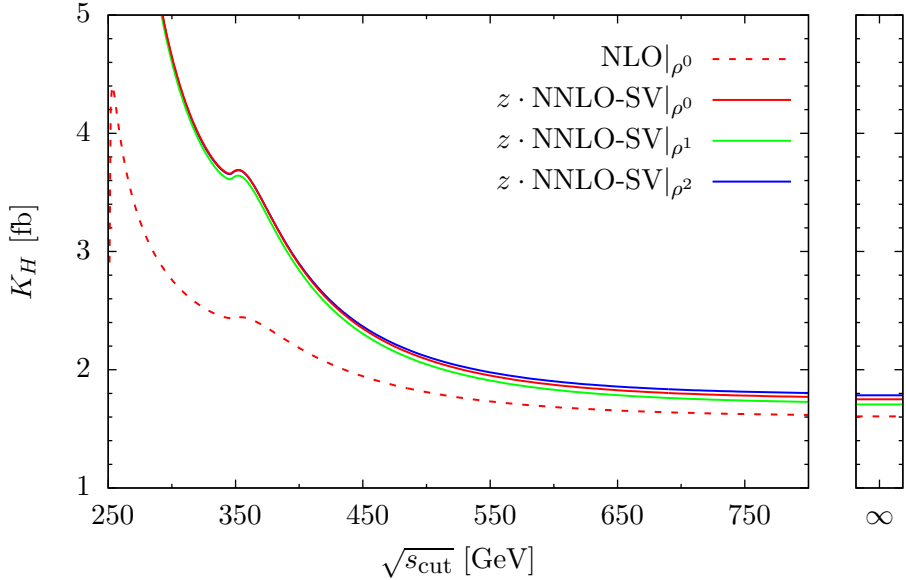}
    \\
    \caption{\label{fig::sig_K_hadr_nnlo} Hadronic NLO (dotted) and NNLO
      (solid) $K$ factor as a function of $\sqrt{s_{\rm cut}}$.}
  \end{center}
\end{figure}

Fig.~\ref{fig::sig_K_hadr_nnlo} shows the hadronic $K$ factor at NNLO
which is defined by
\begin{align}
  K_H^\textrm{NNLO} &= \frac{\left(\sigma_H^\LO + \delta \sigma_H^\NLO+
      \delta \sigma_H^\NNLO\right)|_{\NNLO \textrm{
        pdfs}}}{\sigma_H^\LO|_{\LO\textrm{ pdfs}}}
  \,.
\end{align}
as a function of $\sqrt{s}_{\rm cut}$. For comparison also the NLO result
from Fig.~\ref{fig::sig_hadr} is shown as dotted curve using the
$M_t\to\infty$ result. One observes that the various $\rho$ orders lead to
similar results for $K_H^\textrm{NNLO}$. Furthermore, there is a strong raise close to
threshold which is due to the steeper behaviour of the NNLO cross section as
can be seen in the inlay of Fig.~\ref{fig::sig_part_lo_nlo_nnlo}. For higher
values of $\sqrt{s_{\rm cut}}$, in particular for the total cross section,
$K_H^\textrm{NNLO}$ approaches $1.7-1.8$.

\begin{table}[t]
\centering
\begin{tabular}{l||l|l|c}
 & $\sigma_H$ [fb]& $K^{\textrm{(N)NLO}}$ & $X^\textrm{NNLO}$ [\%] \\\hline\hline
LO & 22.7 & --- &  --- \\\hline
LO+NLO$|_{\rho^0}$  &  36.4 & 1.60 &  --- \\\hline
LO+NLO$|_{\rho^0}$+NNLO$|_{\rho^0}$ &39.7 & 1.75 & 0 \\
LO+NLO$|_{\rho^0}$+NNLO$|_{\rho^1}$ &38.7 & 1.70 & $-2.5$\\
LO+NLO$|_{\rho^0}$+NNLO$|_{\rho^2}$ &40.5 & 1.78 & $+2.0$ \\\hline
\end{tabular}
\caption{\label{tab:nnlo}Total hadronic cross section at LO, NLO and NNLO-SV
  including top quark mass effects using $\mu = 2 m_H$ and $f(z)=z$.
  }
\end{table}

Results for the total cross section at LO, NLO and NNLO 
are summarized in Table~\ref{tab:nnlo} for $\mu=2m_H$. 
At NLO only $\rho^0$ terms are included in the analysis
whereas at NNLO $\rho^0$, $\rho^1$ and $\rho^2$ terms are considered.
In this way we can estimate the top mass effects of the NNLO term.
Besides the cross section also the $K$ factor is shown.
At NNLO we use $f(z)=z$ and 
we furthermore show the relative deviation due to $1/M_t^{2n}$ terms
defined through
\begin{align}
  X^\NNLO &= \frac{\delta\sigma_H^\NNLO|_{\rho^n} -
    \delta\sigma_H^\NNLO|_{\rho^0}}{\sigma_H^{\NNLO}|_{\rho^0}}
  \,.
\end{align}
The two known mass correction terms lead to a change of the cross section
by about $\pm 2\%$. Assuming a similar pattern as at NLO we thus estimate 
that NNLO top quark mass corrections change the effective-theory result by
at most $\pm 5\%$.

%- }}}
%- {{{ Conclusions:

\section{\label{sec::con}Conclusions}

We compute NLO and NNLO corrections to double Higgs boson production in gluon
fusion beyond the effective-theory approach. The starting point of the
calculation are full-theory Feynman diagrams. We perform an asymptotic
expansion in the limit where the top quark mass is large and compute at NNLO 
three terms in the $1/M_t$ expansion for the virtual corrections. They are
used to construct a soft-virtual approximation for the production cross
section. In the limit $M_t\to\infty$ the effective-theory
result of Ref.~\cite{deFlorian:2013uza} is confirmed and $1/M_t^2$
and $1/M_t^4$ terms are added. 

The main result of this paper is illustrated in
Fig.~\ref{fig::sig_hadr_lo_nlo_nnlo} where the hadronic cross section is shown
as a function of $\sqrt{s_{\rm cut}}$ (a technical cut on the partonic
center-of-mass energy). The curves including $1/M_t$ corrections deviate from
the infinite mass result only by a few per cent which leads us to the estimate
that the effective-theory result is accurate to $\pm 5\%$.  Analog
results for the mass corrections at NLO are shown in Fig.~\ref{fig::sig_hadr}
which constitutes an update of Ref.~\cite{Grigo:2013rya}. Here we estimate the
uncertainty to $\pm 10\%$.

We want to stress that the results obtained in this paper provide excellent
approximations for small values of the partonic center-of-mass energy, say
below $\sqrt{s}\approx 400$~GeV. Although in this region the cross section is
small it is of interest since there the cross section has a characteristic
behaviour. Furthermore, it is possible to use our result in this region as a
benchmark for future calculations taking into account the exact dependence on
$M_t$.

The methods described in Section~\ref{sec::nnlo} can also be used to compute
top mass corrections to the real radiation part.  However, the simplifications
used in Ref.~\cite{deFlorian:2013jea} where results have been obtained for
$M_t\to \infty$ do not apply once finite mass effects are considered. The
calculation is much more challenging since significantly more Feynman diagrams
contribute and more complicated master integrals have to be computed.

In this paper for the first time the effect of a finite top quark mass
has been examined for the NNLO cross section for double Higgs boson production.
Whereas at NLO an exact calculation is within reach this is certainly not the
case at NNLO. Thus our results become particular important once our NLO
approximations are compared to an exact calculation which increases the
confidence in the uncertainty estimate. Furthermore, one probably can
obtain a prescription to tune the approximation procedure and hence reduce
the uncertainty at NNLO.

%- }}}
%- {{{ Ackn.:

\section*{Acknowledgments}

We would like to thank Kirill Melnikov for providing to us the analytic
result for $I_1(4)$ in Eq.~(\ref{eq::I14}), for many useful discussions and
for carefully reading the manuscript.
This work is supported by the Deutsche Forschungsgemeinschaft through grant
STE~945/2-1 and by KIT through its distinguished researcher fellowship
program. Parts of this work were supported by the European Commission through
contract PITN-GA-2012-316704 (HIGGSTOOLS).
J.G. would like to express a special thanks to the Mainz Institute for
Theoretical Physics (MITP) for its hospitality and support.

%- }}}

%- {{{ bibliography

%- }}}


\begin{thebibliography}{99}

%
% gghh_sv_ref.tex -- generated by sortref-2.3.5  
% ((C) R. Harlander, http://www.robert-harlander.de/software/)
% on Fri Jul 24 13:07:56 CEST 2015
%

%1
\bibitem{Djouadi:1999rca}
  A.~Djouadi, W.~Kilian, M.~M\"uhlleitner and P.~M.~Zerwas,
  Eur.\ Phys.\ J.\ C {\bf 10} (1999) 45
  [hep-ph/9904287].
  %%CITATION = HEP-PH/9904287;%%

%2
\bibitem{Baur:2002qd}
  U.~Baur, T.~Plehn and D.~L.~Rainwater,
  Phys.\ Rev.\ D {\bf 67} (2003) 033003
  [hep-ph/0211224].
  %%CITATION = HEP-PH/0211224;%%

%3
\bibitem{Baur:2003gp}
  U.~Baur, T.~Plehn and D.~L.~Rainwater,
  Phys.\ Rev.\ D {\bf 69} (2004) 053004
  [hep-ph/0310056].
  %%CITATION = HEP-PH/0310056;%%

%4
\bibitem{Dolan:2012rv}
  M.~J.~Dolan, C.~Englert and M.~Spannowsky,
  JHEP {\bf 1210} (2012) 112
  [arXiv:1206.5001 [hep-ph]].
  %%CITATION = ARXIV:1206.5001;%%

%5
\bibitem{Papaefstathiou:2012qe}
  A.~Papaefstathiou, L.~L.~Yang and J.~Zurita,
  Phys.\ Rev.\ D {\bf 87} (2013) 011301
  [arXiv:1209.1489 [hep-ph]].
  %%CITATION = ARXIV:1209.1489;%%

%6
\bibitem{Baglio:2012np}
  J.~Baglio, A.~Djouadi, R.~Gr\"ober, M.~M.~M\"uhlleitner, J.~Quevillon and
  M.~Spira,
  JHEP {\bf 1304} (2013) 151
  [arXiv:1212.5581 [hep-ph]].
  %%CITATION = ARXIV:1212.5581;%%

%7
\bibitem{Goertz:2013kp}
  F.~Goertz, A.~Papaefstathiou, L.~L.~Yang and J.~Zurita,
  JHEP {\bf 1306} (2013) 016
  [arXiv:1301.3492 [hep-ph]].
  %%CITATION = ARXIV:1301.3492;%%

%8
\bibitem{Gouzevitch:2013qca}
  M.~Gouzevitch, A.~Oliveira, J.~Rojo, R.~Rosenfeld, G.~P.~Salam and V.~Sanz,
  JHEP {\bf 1307} (2013) 148
  [arXiv:1303.6636 [hep-ph]].
  %%CITATION = ARXIV:1303.6636;%%

%9
\bibitem{Glover:1987nx}
  E.~W.~N.~Glover and J.~J.~van der Bij,
  Nucl.\ Phys.\ B {\bf 309} (1988) 282.
  %%CITATION = NUPHA,B309,282;%%

%10
\bibitem{Plehn:1996wb}
  T.~Plehn, M.~Spira and P.~M.~Zerwas,
  Nucl.\ Phys.\ B {\bf 479} (1996) 46
   [Erratum-ibid.\ B {\bf 531} (1998) 655]
  [hep-ph/9603205].
  %%CITATION = HEP-PH/9603205;%%

%11
\bibitem{Dawson:1998py} 
  S.~Dawson, S.~Dittmaier and M.~Spira,
  Phys.\ Rev.\ D {\bf 58}, 115012 (1998)
  [hep-ph/9805244].
  %%CITATION = HEP-PH/9805244;%%

%12
\bibitem{Grigo:2013rya}
  J.~Grigo, J.~Hoff, K.~Melnikov and M.~Steinhauser,
  Nucl.\ Phys.\ B {\bf 875} (2013) 1
  [arXiv:1305.7340 [hep-ph]].
  %%CITATION = ARXIV:1305.7340;%%

%13
\bibitem{Maltoni:2014eza}
  F.~Maltoni, E.~Vryonidou and M.~Zaro,
  JHEP {\bf 1411} (2014) 079
  [arXiv:1408.6542 [hep-ph]].
  %%CITATION = ARXIV:1408.6542;%%

%14
\bibitem{deFlorian:2013uza}
  D.~de Florian and J.~Mazzitelli,
  Phys.\ Lett.\ B {\bf 724} (2013) 306
  [arXiv:1305.5206 [hep-ph]].
  %%CITATION = ARXIV:1305.5206;%%

%15
\bibitem{deFlorian:2013jea}
  D.~de Florian and J.~Mazzitelli,
  Phys.\  Rev.\  Lett.\  111, {\bf 201801} (2013)
  [arXiv:1309.6594 [hep-ph]].
  %%CITATION = ARXIV:1309.6594;%%

%16
\bibitem{Grigo:2014jma}
  J.~Grigo, K.~Melnikov and M.~Steinhauser,
  Nucl.\ Phys.\ B {\bf 888} (2014) 17
  [arXiv:1408.2422 [hep-ph]].
  %%CITATION = ARXIV:1408.2422;%%

%17
\bibitem{Shao:2013bz}
  D.~Y.~Shao, C.~S.~Li, H.~T.~Li and J.~Wang,
  arXiv:1301.1245 [hep-ph].
  %%CITATION = ARXIV:1301.1245;%%

%18
\bibitem{deFlorian:2015moa}
  D.~de Florian and J.~Mazzitelli,
  arXiv:1505.07122 [hep-ph].
  %%CITATION = ARXIV:1505.07122;%%

%19
\bibitem{Anastasiou:2002yz} 
  C.~Anastasiou and K.~Melnikov,
  Nucl.\ Phys.\ B {\bf 646}, 220 (2002)
  [hep-ph/0207004].
  %%CITATION = HEP-PH/0207004;%%

%20
\bibitem{Grigo:2013xya}
  J.~Grigo, J.~Hoff, K.~Melnikov and M.~Steinhauser,
  PoS RADCOR {\bf 2013} (2013) 006
  [arXiv:1311.7425 [hep-ph]].
  %%CITATION = ARXIV:1311.7425;%%

%21
\bibitem{Grigo:2014oqa}
  J.~Grigo and J.~Hoff,
  arXiv:1407.1617 [hep-ph].
  %%CITATION = ARXIV:1407.1617;%%

%22
\bibitem{deFlorian:2012za}
  D.~de Florian and J.~Mazzitelli,
  JHEP {\bf 1212} (2012) 088
  [arXiv:1209.0673 [hep-ph]].
  %%CITATION = ARXIV:1209.0673;%%

%23
\bibitem{Catani:1998bh}
  S.~Catani,
  Phys.\ Lett.\ B {\bf 427} (1998) 161
  [hep-ph/9802439].
  %%CITATION = HEP-PH/9802439;%%

%24
\bibitem{Martin:2009iq}
  A.~D.~Martin, W.~J.~Stirling, R.~S.~Thorne and G.~Watt,
  Eur.\ Phys.\ J.\ C {\bf 63} (2009) 189
  [arXiv:0901.0002 [hep-ph]].
  %%CITATION = ARXIV:0901.0002;%%

%25
\bibitem{Agashe:2014kda}
  K.~A.~Olive {\it et al.}  [Particle Data Group Collaboration],
  Chin.\ Phys.\ C {\bf 38} (2014) 090001.
  %%CITATION = CHPHD,C38,090001;%%

%26
\bibitem{Aad:2015zhl}
  G.~Aad {\it et al.} [ATLAS and CMS Collaborations],
  Phys.\ Rev.\ Lett.\  {\bf 114} (2015) 191803
  [arXiv:1503.07589 [hep-ex]].
  %%CITATION = ARXIV:1503.07589;%%

%27
\bibitem{Anastasiou:2014vaa}
  C.~Anastasiou, C.~Duhr, F.~Dulat, E.~Furlan, T.~Gehrmann, F.~Herzog
  and B.~Mistlberger,
  Phys.\ Lett.\ B {\bf 737} (2014) 325
  [arXiv:1403.4616 [hep-ph]].
  %%CITATION = ARXIV:1403.4616;%%

%28
\bibitem{Herzog:2014wja}
  F.~Herzog and B.~Mistlberger,
  arXiv:1405.5685 [hep-ph].
  %%CITATION = ARXIV:1405.5685;%%

%29
\bibitem{Nogueira:1991ex}
  P.~Nogueira,
  J.\ Comput.\ Phys.\  {\bf 105} (1993) 279.
  %%CITATION = JCTPA,105,279;%%

%30
\bibitem{Kuipers:2012rf}
  J.~Kuipers, T.~Ueda, J.~A.~M.~Vermaseren and J.~Vollinga,
  Comput.\ Phys.\ Commun.\  {\bf 184} (2013) 1453
  [arXiv:1203.6543 [cs.SC]].
  %%CITATION = ARXIV:1203.6543;%%

%31
\bibitem{Harlander:1997zb}
  R.~Harlander, T.~Seidensticker and M.~Steinhauser,
  Phys.\ Lett.\ B {\bf 426} (1998) 125
  [hep-ph/9712228].
  %%CITATION = HEP-PH/9712228;%%

%32
\bibitem{Seidensticker:1999bb}
  T.~Seidensticker,
  hep-ph/9905298.
  %%CITATION = HEP-PH/9905298;%%

%33
\bibitem{Steinhauser:2000ry}
  M.~Steinhauser,
  Comput.\ Phys.\ Commun.\  {\bf 134} (2001) 335
  [hep-ph/0009029].
  %%CITATION = HEP-PH/0009029;%%

%34
\bibitem{Smirnov:2013dia}
  A.~V.~Smirnov and V.~A.~Smirnov,
  arXiv:1302.5885 [hep-ph].
  %%CITATION = ARXIV:1302.5885;%%

%35
\bibitem{Smirnov:2014hma}
  A.~V.~Smirnov,
  Comput.\ Phys.\ Commun.\  {\bf 189} (2014) 182
  [arXiv:1408.2372 [hep-ph]].
  %%CITATION = ARXIV:1408.2372;%%

%36
\bibitem{Birthwright:2004kk}
  T.~G.~Birthwright, E.~W.~N.~Glover and P.~Marquard,
  JHEP {\bf 0409} (2004) 042
  [hep-ph/0407343].
  %%CITATION = HEP-PH/0407343;%%

%37
\bibitem{Ellis:2007qk}
  R.~K.~Ellis and G.~Zanderighi,
  JHEP {\bf 0802} (2008) 002
  [arXiv:0712.1851 [hep-ph]].
  %%CITATION = ARXIV:0712.1851;%%

%38
\bibitem{Chavez:2012kn}
  F.~Chavez and C.~Duhr,
  JHEP {\bf 1211} (2012) 114
  [arXiv:1209.2722 [hep-ph]].
  %%CITATION = ARXIV:1209.2722;%%

%39
\bibitem{Goncharov:1998kja}
  A.~B.~Goncharov,
  Math.\ Res.\ Lett.\  {\bf 5} (1998) 497
  [arXiv:1105.2076 [math.AG]].
  %%CITATION = ARXIV:1105.2076;%%

%40
\bibitem{Smirnov:2013eza}
  A.~V.~Smirnov,
  Comput.\ Phys.\ Commun.\  {\bf 185} (2014) 2090
  [arXiv:1312.3186 [hep-ph]].
  %%CITATION = ARXIV:1312.3186;%%

%41
\bibitem{diss_Hoff}
  J. Hoff, ``Methods for multiloop calculations and Higgs boson production at
  the LHC'', Dissertation, KIT, 2015.

%42
\bibitem{Vermaseren:2000nd}
  J.~A.~M.~Vermaseren,
  math-ph/0010025.
  %%CITATION = MATH-PH/0010025;%%

%43
\bibitem{Gehrmann:2005pd}
  T.~Gehrmann, T.~Huber and D.~Maitre,
  Phys.\ Lett.\ B {\bf 622} (2005) 295
  [hep-ph/0507061].
  %%CITATION = HEP-PH/0507061;%%

%44
\bibitem{Pak:2011hs}
  A.~Pak, M.~Rogal and M.~Steinhauser,
  JHEP {\bf 1109} (2011) 088
  [arXiv:1107.3391 [hep-ph]].
  %%CITATION = ARXIV:1107.3391;%%


\end{thebibliography}
\end{document}